\documentclass[letterpaper,11pt]{article}
\pdfoutput=1 

\usepackage{jheppub} 

\usepackage{tikz}
\usepackage{dsfont}
\usepackage{amsmath}
\def\ba{\begin{align}}\def\ea{\end{align}}

\definecolor{darkgreen}{rgb}{0,0.4,0}
\def\beq{\begin{eqnarray}}\def\eeq{\end{eqnarray}}
\def\be{\begin{equation}}\def\ee{\end{equation}}
\def\ben{\begin{equation}}
\def\een{\end{equation}}
\def\bea{\begin{eqnarray}}
\def\eea{\end{eqnarray}}

\def\vx{{\vec{x}}}

\def\cO{{\cal{O}}}

\def\bL{{\bar L}}

\def\t6 {T_\mt{D6}}

\newcommand{\mt}[1]{\textrm{\tiny #1}}

\def\cale         {{\cal E}}

\def\mI{\mathbb{I}}

\def\ee           {{\rm e}}

\def\sqr#1#2{{\vcenter{\vbox{\hrule height.#2pt
 \hbox{\vrule width.#2pt height#1pt \kern#1pt
 \vrule width.#2pt}\hrule height.#2pt}}}}

\def\ee{\cale}

\def\aa1{\phi}
\def\cc1{\psi}

\def\vx{\vec{x}}

\def\ie{{\it i.e.}}

\title{Dynamical Phases of Higher Dimensional Floquet CFTs}

\author{Diptarka Das $^1$,}
\author{Sumit R. Das$^2$,}
\author{Arnab Kundu$^{3,4}$,}
\author{Krishnendu Sengupta $^5$,}

\affiliation{$^1$ Department of Physics, Indian Institute of Technology, Kanpur, UP 208016, INDIA.}
\affiliation{$^2$Department of Physics and Astronomy, University of Kentucky, Lexington, KY 40506, U.S.A.}
\affiliation{$^3$Saha Institute of Nuclear Physics, 1/AF Bidhannagar, Kolkata 700064, INDIA.}
\affiliation{$^4$Homi Bhaba National Institute, Training School Complex, Anushaktinagar, Mumbai 400094, INDIA.}
\affiliation{$^5$School of Physical Sciences, Indian Association for the Cultivation of Science, Jadavpur, Kolkata 700032, INDIA.}

\emailAdd{didas@iitk.ac.in}
\emailAdd{das@pa.uky.edu}
\emailAdd{arnab.kundu@saha.ac.in}
\emailAdd{tpks@iacs.res.in}

\abstract{
This paper investigates the dynamical phases of Floquet Conformal Field Theories (CFTs) in space-time dimensions greater than two. Building upon our previous work \cite{ddks1} which introduced quaternionic representations for studying Floquet dynamics in higher-dimensional CFTs, we now explore more general square pulse drive protocols that go beyond a single $SU(1,1)$ subgroup. We demonstrate that, for multi-step drive protocols, the system exhibits distinct dynamical phases characterized  by the nature of the eigenvalues of the quaternionic matrix representing time evolution in a single cycle, leading to different stroboscopic responses. Our analysis establishes a fundamental geometric interpretation where these dynamical phases directly correspond to the presence or absence of Killing horizons in the base space  of the CFT and in a higher dimensional AdS space on which a putative dual lives. The heating phase is associated with a non-extremal horizon, the critical phase with an extremal horizon which disappears in the non-heating phase. We develop perturbative approaches to compute the Floquet Hamiltonians in different regimes and show, how tuning drive parameters can lead to horizons, providing a geometric framework for understanding heating phenomena in driven conformal systems. 
}

\begin{document}

\begin{flushright}
\end{flushright}

\maketitle
\flushbottom
\vspace{10pt}

\section{Introduction}

The physics of quantum systems driven out of equilibrium has received a lot of attention in recent times \cite{rev1,rev2,rev3,rev4,rev5,rev6,rev7,rev8,rev9,rev10,rev11,rev12}. A class of such systems constitutes quantum systems subjected to time-periodic drive characterized by a time period $T=2\pi/\omega_D$ where $\omega_D$ is the drive frequency. Such periodically driven systems exhibit phases that are not possible in equilibrium. These include dynamical localization caused by suppression of tunnelling, suppression of ionization, dynamical freezing,  and time crystalline phases of quantum matter \cite{rev5,rev6,rev7,rev8,rev9,rev10,rev11,rev12}. Many of these phenomena have their roots in an approximate emergent symmetry of the Floquet Hamiltonian that describes these systems \cite{rev12}; such an emergence has no analogue either in equilibrium or for quantum systems subjected to aperiodic drives. 

While these phenomena are of great theoretical and experimental interest, the analytical tools available to understand such dynamics are rather limited \cite{rev11}. This aspect may be traced back to the difficulty in {\color{black} the} computation of the Floquet Hamiltonian $H_F$, which describes the time evolution over a single time period $T$ (see below). Known analytic methods for computing $H_F$ include a high frequency expansion (Magnus expansion) and Floquet perturbation theory (FPT), which is accurate in the high drive amplitude regime. Both these methods require a perturbative parameter such as the inverse of either the drive frequency (Magnus expansion) or the drive amplitude (FPT). In other parameter regimes, where the drive frequency or amplitude is not high, an investigation of such systems relies on numerical calculations.

More recently, it has been realized that driven $1+1$ dimensional systems at criticality can be solved using the tools of conformal field theory \cite{wen0,wen1,fan2, chitra1,wen2,Lapierre4,fan1, chitra2,  das1, wen3, nozaki1, Lapierre3, Lapierre2, wen4, suchetan1, nozaki2, nozaki3, deBoer, nozaki4, nozaki5, mezei,Lapierre1, nozaki6, nozaki7,  arnab2}. These systems can be alternatively thought of as conformal field theories on a curved space-time \cite{wen0,maccormack,deBoer,caputa, caputa2}. The simplest examples of periodic drives in these systems are piecewise constant Hamiltonians (a square pulse protocol): the Hamiltonians in different pieces are different linear combinations of the conformal generators that belong to an $SU(1,1)$ subalgebra of the Virasoro algebra. Representative simple examples are $L_0 + \bL_0, L_{\pm 1} + \bL_{\pm 1}$. Since the time-evolution in a single step is a conformal transformation, in which the Lorentzian time appears as a parameter, physical quantities can be computed using conformal transformations. In these systems, it was found that by changing the time extent of each of these intervals, the systems can be driven between different dynamical phases. These phases are characterized by the type of conjugacy class of the $SU(1,1)$ transformation for a single time period which determines the properties of the operator for time evolution over time $T$.

When the type of conjugacy class is hyperbolic, the stroboscopic response is exponential in the number of time periods, $n$. This is called a heating phase - unequal time correlation functions decay as a function of $n$ for large $n$ in this phase. Even though this resembles a thermal phase, the system is not featureless. For a CFT defined on a circle, the conformal transformation which corresponds to time evolution over a single period has two fixed points: the entanglement entropy of an interval which contains such a fixed point grows linearly, while it quickly goes to zero if the interval does not contain a fixed point. Likewise, the energy density gets peaked near these fixed point. In contrast, when $U(T)$ is in the elliptic type of conjugacy class, there are no fixed points on the circle: now the trajectory of a point on the circle under stroboscopic evolution oscillates, leading to oscillatory behavior of all physical quantities. When the type of conjugacy class is parabolic, the behaviour of various quantities is a power law in $n$, which we call a critical phase. Since the type of conjugacy classes depend on the time steps within each period, it is possible to change the dynamical phase by changing the time steps.

$1+1$ dimensional conformal field theories are of course very special and one might wonder whether any of these interesting features hold for higher-dimensional field theories as well. In \cite{ddks1} we initiated a study of such Floquet conformal field theories in $d+1$ dimensions defined on $S^d \times$~~(time), for $d=2,3$. The main technical tool in this study is a representation of conformal transformations in these dimensions in terms of Mobius transformations of {\em quaternions}. We studied square pulse protocols with time steps of extent $T_i$, in which the Hamiltonians are constant and linear combinations of generators of $SU(1,1)$ sub-algebras of the full conformal algebra. In this work it was shown that under the time-evolution with a {\em single} Hamiltonian, the system can be in one of three phases, depending on a parameter $\beta$, which determines the amplitude of the deformation. For $\beta < 1$, various physical observables such as the stroboscopic time evolution  of the unequal time correlators starting from the initial vacuum state and that of fidelity and energy density of a primary state exhibit oscillations with a time period $T_p = 1/\sqrt{1-\beta^2}$. For $\beta > 1$ these decay exponentially in time, while for $\beta = 1$, the response is {\color{ black} power law} in time. These three {\em static} phases correspond to the three types of conjugacy classes of the $SU(1,1)$ transformation that acts as the time evolution. 

These static phases should be contrasted with the stroboscopic response in a periodic drive with a multi-step square pulse protocol, with different Hamiltonians in different steps. In such a drive we are interested in the behavior of various quantities measured at the end of $n$ time periods, as a function of $n$.  Restricting to Hamiltonians which belong to a {\em single} $SU(1,1)$ sub-algebra, we found that the stroboscopic response is very similar to that in $1+1$ dimensions described above. This is expected since the Floquet Hamiltonian now belongs to the same $SU(1,1)$ subalgebra.

In this paper, we extend our discussion to processes whose Floquet Hamiltonians belong to more general elements of the full conformal group and obtain geometric  interpretations of the results. We find a variety of dynamical phases as measured by the stroboscopic response, determined by the amplitudes and time interval of the individual steps. As will be discussed below, this response is governed by a set of functions of the number of steps $n$ (and the parameters in the Hamiltonians) which determine how a point on the $S^d \times$~~(time) moves under time evolution. As in $1+1$ dimensions we find phases in which the $n$ dependence of all these functions can be purely oscillatory, power law or exponential. However for $d > 1$ we find in addition a phase where some of these functions are oscillatory, while others are exponential. 

We show that the static phases for a single Hamiltonian, or the dynamical phases in a periodic drive resulting {\color{ black} in} a Floquet Hamiltonian describing the stroboscopic response, have natural geometric interpretations. In the CFT, a given Hamiltonian or a Floquet Hamiltonian defines a conformal Killing vector on the base space. We find that this conformal Killing vector remains time-like everywhere in an oscillatory phase. In a heating phase, this ceases to be the case and there are Killing horizons characterized by a non-zero surface gravity. In the critical phase, the Killing horizon becomes extremal with a vanishing surface gravity. This means that the heating observed can be thought of as a generalized Unruh effect. This is similar to the situation in $1+1$ dimensional quench from a standard Hamiltonian (dilatation) to an SSD Hamiltonian, or a periodic drive with alternative standard and SSD Hamiltonians where there are extremal horizons \cite{mezei}. For holographic $d+1$ dimensional CFT's which have a bulk gravity dual in asymptotically $AdS_{d+2}$ we show that these horizons lift into horizons in the bulk.

The plan of the rest of the paper is as follows. In Sec.\ \ref{setup} we describe the basic setup and briefly summarize our results. In Sec.\ \ref{dk} we provide examples of deformed Hamiltonians and study the corresponding Killing horizons. This is followed by Sec.\ \ref{ctrans}, where we provide details of the quaternion formalism used to study the dynamics of driven CFTs. Next, we study {\color{ black} the} dynamics of periodically driven CFTS in Sec.\ \ref{fldyn}. This is followed by a study of perturbative Floquet Hamiltonians in Sec.\ \ref{pert}; the associated bulk Killing horizons are analyzed in Sec.\ \ref{killing1}.  Finally, we discuss our main results and conclude in Sec.\ \ref{diss}. Some details of the calculations are presented in the appendices.

\section{The Set-up and a Summary of results}
\label{setup}

{\color {black} In this section, we chart out the basic setup and provide a context for our work. This is followed by a detailed summary of the main results of this work. 

In what follows,} {\color{ black} we will consider $d+1$ dimensional conformal field theories on (time) $ \times S^d$ for $d=2,3$. By a choice of units we set the radius of $S^d$ to be unity}. We will consider a class of Hamiltonians given by 
\ben
H_\mu(\beta_\mu) = \int d^d \Omega [ 1 + \beta_\mu Y^\mu ] T_{tt}~~~~~~{\rm (no~ sum~ over }~~~\mu ) \ .
\label{1-2}
\een
The integral is over a unit sphere $S^d$ described in terms of a unit vector $Y^\mu$ in $R^{d+1}$ by $\sum_\mu Y_\mu Y^\mu = 1$.
$\beta_\mu$ is a real parameter and $T_{tt}$ is the energy density. In terms of the generators of the conformal algebra the Hamiltonian $H_\mu (\beta_\mu)$ is
\ben
H_\mu (\beta_\mu) = 2iD + i\, \beta_\mu (K_\mu + P_\mu) ~~~~~~~~~~{\rm (no~ sum~ over }~~~\mu )
\label{1-3}
\een
where $D, K_\mu, P_\mu$ are the dilatation, special conformal transformation and momentum generators of the conformal algebra. (Our conventions are described in Appendix A.)
A given $H_\mu (\beta_\mu)$ belongs to an $SU(1,1)$ sub-algebra of the full $SO(d+1,2)$ conformal algebra. 

In this paper we will consider the Heisenberg picture state to be the conformally invariant vacuum. This state remains unchanged during the time evolution. However, the spectra of the Hamiltonians $H_\mu(\beta_\mu)$ are different. As a result, unequal time vacuum correlation functions are different for different Hamiltonians. Other diagnostics are fidelities and expectation values of operators in excited primary states. In \cite{ddks1} it was shown that under time evolution by a given Hamiltonian $H_\mu(\beta_\mu)$ these responses display exponential decay in time for $\beta_\mu > 1$, power law decay for $\beta_\mu = 1$ and purely oscillatory behavior for $\beta_\mu < 1$. 

The periodic drives we use are square pulse protocols. Each time period $T$ of the drive consists of several steps with a different Hamiltonian for each step. The stroboscopic response is determined by the Floquet Hamiltonian $H_F(T)$ which describes time evolution in a single period,
{\color{ black}
\ben
U(T,0) \equiv U(T)= {\mathcal T} \exp \left[-i \int_0^T dt H(t) /\hbar\right] = \exp[-i H_F(T) T/\hbar]
\label{1-1}
\een}
$H_F(T)$ depends on the extents of the individual time steps as well as the $\beta_\mu$'s.

Consider for example a periodic two step square pulse protocol with a time period $T=T_1+T_2$ would be (with $T_1 < T$)
\bea
H & = & 2iD~~~~~~~~~~~~~{\rm for} ~~~nT \leq t \leq nT + T_1 \nonumber \\
H & = & H_0 (\beta_0)~~~~~~~~~{\rm for}~~~  nT + T_1 \leq t \leq (n+1)T
\label{1-4}
\eea
with $n=0,1,2,\cdots$. The Floquet Hamiltonian now belongs to the $SU(1,1)$ subalgebra formed by $D,K_0,P_0$ of the full conformal algebra. 
In \cite{ddks1}, we showed that {\em regardless of the value $\beta_0$}, the stroboscopic response can correspond to a heating phase in which the response after some time $nT$ is an exponential function of $n$, a critical phase where the dependence is a power law in $n$, or a non-heating {\color{ black} phase where the response oscillates as a function of $n$}, depending on the values of $T_1, T$. In particular, consider the case $\beta_0 < 1$ so that the Hamiltonians in both the steps are individually in non-heating phases with oscillatory response. However, by tuning the values of $T_1, T_2$ one can drive the system to a heating phase. Similarly, there are non-heating phases for stroboscopic response even for $\beta_0 > 1$. These phases are therefore {\em dynamical} and follow from the explicit form of the Floquet Hamiltonian $H_F$ which is now a linear combination of the generators $D,K_0,P_0$. The phases are once again characterized by the type of conjugacy class of the $SU(1,1)$ evolution operator for a single period $U(T) = e^{-iH_FT}$.

The simplest square pulse protocol which goes beyond a $SU(1,1)$ is a two step process
\bea
H & = & H_1 (\beta_1)~~~~~~~~~~~~~{\rm for} ~~~nT \leq t \leq nT + T_1 \nonumber \\
H & = & H_0 (\beta_0)~~~~~~~~~{\rm for}~~~  nT + T_1 \leq t \leq (n+1)T
\label{1-5}
\eea
The generators involved now are $D,K_1,K_0,P_1,P_0$ which would lead to a Floquet Hamiltonian which is a linear combination of all generators of the full conformal algebra in $1+1$ dimensions, $SO(2,2)$. Adding a 3rd step which involves the Hamiltonian $H_2 (\beta_2)$ enlarges this to the conformal group $SO(3,2)$ in $2+1$ dimensions, and so on. In \cite{ddks1} we considered protocols like (\ref{1-5}) and obtained the fidelity of a primary state at the end of one cycle, but did not examine the dynamical phases which result from a periodic drive; {\color {black} analyzing such dynamical phases is one of the key aspects of the present work.

To this end,} as in \cite{ddks1}, we will represent the coordinates of the base space of the field theory in terms of quaternions \footnote{In Lorentzian signature these are in fact special bi-quaternions. However we will simply refer them to quaternions.}, which are adequate to describe $d+1$ dimensions with $d=1,2,3$. A general conformal transformation then acts on these quaternions via a $2 \times 2$ matrix with quaternionic entries, a quaternionic Mobius transformation. In Euclidean signature this is an element of $SO(2,H)$. Time evolution of the kind described above is such a conformal transformation whose parameter is the time itself. 

An efficient way to determine the stroboscopic response is to determine the eigenvalues and eigenvectors of the matrix which represents real time evolution in a single time period. As will be discussed in detail (Section 5) this will allow us to determine the conformal transformation representing the time evolution after $n$ periods easily. The response is then encoded in the dependence of the $4 \times 4$ quaternion matrix on $n$. When all the eigenvalues are pure phases, the matrix elements of the quaternion matrix are periodic functions of $n$ and this leads to oscillatory behavior of the unequal time correlator as a function of $n$. When the eigenvalues are all real these depend on $n$ exponentially, leading to exponential decay of the correlator as a function of $n$, which is a dynamical phase resembling heating. When half of the eigenvalues are real and the other half pure phase, some of the functions grow exponentially, while others oscillate - we will call this a ``hybrid phase'' {\color{ black} consisting of oscillations on top of an exponential decay}. Such a hybrid phase does not exist for $SU(1,1)$ drives {\color{black} involving just one direction}. Finally there can be a critical dynamical phase where the correlators decay as a power law in $n$. 

For a 2-step process of the type (\ref{1-5}) and for a 3 step process where one of the $\beta$'s vanish (i.e. the time evolution is by $D$ in one of the steps) we can determine the dynamical phases analytically, and we find all the four phases described above. For a 3-step process with the three steps involving $K_\mu + P_\mu$ in three different directions, we can determine the eigenvalues numerically. Up to numerical uncertainties we do not find a purely oscillatory phase. This leads us to speculate that in a generic drive an oscillatory phase does not exist. {\color{ black} This is the main distinguishing feature of higher dimensional Floquet CFT's.}

As in $1+1$ dimensions, the phases should be characterized by the types of conjugacy classes of the corresponding group element $SO(3,2)$ or $SO(4,2)$ which are in turn determined by the eigenvalues of the corresponding quaternionic matrix. It will be interesting to see if the types of conjugacy classes of $SL(2,H)$ {\color{ black}(the group for conformal transformations in Euclidean space )} discussed in \cite{parker-short} can be extended to types of conjugacy classes of bi-quaternionic Mobius transformations appropriate for real time evolution \footnote{The types of conjugacy classes of both $SU(1,1)$ and $SL(2,R)$ are determined by the value of the trace of the corresponding $2 X 2$ representation. One needs the analogous results for (bi)-quaternionic Mobius transformations.}

When the unitary operator at the end of a single cycle belongs to $SU(1,1)$, the Floquet Hamiltonian has been obtained in \cite{ddks1} - this is possible by resorting to the Pauli matrix representations for the generators. For more general drives, such an exact calculation is not yet possible.
We show, however, that one can resort to their approximate forms in the high-frequency and high-amplitude regimes. These regimes allow for analytic, albeit perturbative, computation of the Floquet Hamiltonians, using Magnus expansion (high-frequency regime) or FPT (high amplitude regime). This has been done for the two step and the three step processes where the drives go beyond $SU(1,1)$ subgroups. An exact calculation of the Floquet Hamiltonian should be possible using a higher dimensional Clifford algebra representation of the generators; we leave this as a possible subject of future work.

 Our strategy holds for Floquet Hamiltonians which belong to the full (global) conformal algebra in a general $(d+1)$-dimensional system, or its appropriate sub-algebra. Our framework and results are deeply rooted in the symmetry structure of the system, and not on the details of the CFT. This is because we have not used explicitly any OPE data in our calculations, which dictate three and higher point correlation functions in the corresponding CFT and are intrinsically dynamical in nature. While we expect that a much richer dynamical structure is encoded in higher point correlation functions, see {\it e.g.}~\cite{suchetan1}, our analyses here will certainly provide the necessary starting point for them.

As mentioned in the Introduction, the static phases of the Hamiltonians $H_\mu(\beta_\mu)$ and the dynamical phases of the Floquet Hamiltonian are associated with the presence or absence of Killing horizons on the base space-time on which the CFT is defined. For holographic CFT's these horizons extend into the bulk gravitational dual. More precisely, for a $d+1$ dimensional CFT, in its invariant ground state, the bulk geometry is $AdS_{d+2}$: at the boundary, the bulk Killing vectors become the CFT conformal Killing vectors. To better and naturally address such holographic CFTs in general dimensions, we provide here parallel analyses which are  based on Killing vectors in an AdS$_{d+2}$ geometry. 

{\color {black} It is well-known that} the primary role of this AdS-background is to provide a geometry whose isometry group matches with the conformal symmetry group of the CFT. Correspondingly, the CFT Hamiltonian has a natural geometric representation in terms of a corresponding Killing vector in AdS. The choice of the CFT Hamiltonian is, therefore, tied to choosing a particular linear combination of the Killing vectors in AdS. Evidently, the specific CFT coordinates will correspond to the specific patch in AdS. We find the existence of bulk horizons for the heating phases of the driven CFTs, which become extremal at the critical (parabolic) line; no such horizon exists for the non-heating phases. Since one can tune between these phases by tuning the amplitude or frequency of the drive, {\color {black} our work indicates the possibility of tuning between a horizon-less geometry to the one with a horizon in the presence of a periodic drive}.\footnote{ Note that this behaviour of the Killing horizon is strikingly similar to that of a charged black hole with a fixed mass, in which two non-extremal horizons merge into an extremal one and move into the complex-plane, as the charge of the black hole is increased. In this case, the horizon-less configuration yields a naked singularity, which is analogous to the non-heating phase. Of course, in our geometric description, there is no singularity.} {\color{black} It is worth emphasizing that while the holographic description holds literally true for a certain class of CFTs, {\it e.g.}~the ${\mathcal N}=4$ SYM theory in the large $N$ limit and at strong coupling, our discussion here is based completely on the symmetry of the AdS-space. Therefore, the Killing horizon description provides us with a kinematic interpretation for the corresponding horizon dynamics which is in one-to-one correspondence with the CFT dynamics. For ${\mathcal N}=4$ SYM and its close cousins, on the other hand, the actual dynamical evolution will likely have a rich interplay between the kinematic description and the ingredients of the true dynamical duality. Our work here takes the first step towards addressing these issues in future. }

\section{The Deformed Hamiltonians and Killing Horizons}
\label{dk} 

In this section we discuss the three different classes of Hamiltonians $H_\mu(\beta_\mu)$ for a given $\mu$ and the geometric significance for the corresponding vector fields on (time) $\times S^d$. Our conventions for the conformal algebra is given in the Appendix (\ref{algebra}).

\subsection{Hamiltonian Classes}

The nature of time evolution by the Hamiltonian (\ref{1-2}, \ref{1-3}) are very different for $\beta_\mu > 1, \beta_\mu =1$ and $\beta_\mu < 1$. For $\beta_\mu < 1$ define 
\ben
H^\prime_\mu (\beta_\mu) = \frac{1}{\sqrt{1-\beta_\mu^2}} H_\mu (\beta_\mu) = \cosh \chi_\mu (2iD) + i \sinh \chi_\mu (K_\mu + P_\mu)~~~~~~~\beta_\mu = \tanh \chi_\mu
\label{3-1}
\een
Using the conformal algebra it then follows that this Hamiltonian is unitarily equivalent to the standard Hamiltonian of radial quantization 
{\color{ black}
\ben
U_\mu^\dagger H^\prime_\mu (\beta_\mu) U_\mu = 2iD~~~~~~~~
U_\mu = {\rm exp} \left[ - i\frac{\chi_\mu}{2} (K_\mu - P_\mu) \right]
\label{3-2}
\een}
This shows that time evolution using the Hamiltonian (\ref{1-3}) is similar to the standard Hamiltonian. On the other hand when $\beta_\mu > 1$ the redefined Hamiltonian
\ben
H^{\prime \prime}_\mu (\beta_\mu) = \frac{1}{\sqrt{\beta_\mu^2 - 1}} H_\mu (\beta_\mu) = \sinh \xi_\mu (2iD) + i \cosh \xi_\mu (K_\mu + P_\mu)~~~~~~~\beta_\mu = \coth \xi_\mu
\label{3-3}
\een
is unitarily equivalent to the Luscher-Mack Hamiltonian $i(K_\mu + P_\mu)$.
{\color{ black}
\ben
U_\mu^\dagger H^{\prime \prime}_\mu (\beta_\mu) U_\mu = i(K_\mu + P_\mu) ~~~~~~~~
U_\mu = {\rm exp} \left[ - i\frac{\xi_\mu}{2} (K_\mu - P_\mu) \right]
\label{3-3-a}
\een}
Finally when $\beta_\mu = 1$ the Hamiltonian cannot be transformed into anything simpler.

\subsection{Killing Horizons}

A Hamiltonian $H_\mu (\beta_\mu)$ defines a conformal Killing vector field on the base space of the conformal field theory. We will now prove that the three classes of Hamiltonians with $\beta \lesseqgtr 1$ are in one to one correspondence with the presence or absence of Killing horizons in the base space-time of the CFT. For two dimensional sine-squared deformed Hamiltonians such horizons have been studied in \cite{mezei}.
Consider $H_0 (\beta_0)$. Acting on $R^{d+1}$, this is a vector field
\ben
\frac{1}{2}H_0(\beta_0) = \chi= x^\mu \partial_\mu + \frac{\beta}{2}(\partial_{x^0} + 2x^0 x^\mu\partial_\mu -r^2 \partial_{x^0})~~~~~~r^2=x^\mu x_\mu
\label{3-4}
\een
The vector field is therefore
\ben
\chi^\mu = x^\mu + \frac{\beta}{2} (\delta^\mu_0 + 2x^0 x^\mu - r^2\delta^\mu_0)
\label{3-5}
\een
The norm of this vector field is 
\bea
\chi^\mu \chi_\mu & = &  r^2 + \beta_0 (1+r^2)x^0 + \frac{\beta_0^2}{4} (1+r^4 + 4(x^0)^2 - 2r^2) \nonumber \\
& = & \left[re^{i\theta} - \frac{\beta_0}{2} (r^2 e^{2i\theta} +1) \right]
\left[re^{-i\theta} - \frac{\beta_0}{2} (r^2 e^{-2i\theta} +1) \right]
\label{3-6}
\eea
where we have parametrized 
\ben
x^0 = r \cos \theta
\label{3-7}
\een
in anticipation of a Weyl transformation given by (\ref{2-5}).

We need to determine whether this vector is time-like or space-like {\em after} a Weyl transformation followed by an analytic continuation. This means we need to have
\ben
r = e^{it}
\label{3-8}
\een
for real $t$. Because of the Weyl factor (\ref{3-8}) the norm of the vector in real (time) $\times S^d$ is given by ${\cal N} = -e^{-2it} ( \chi^\mu \chi_\mu )$ (the negative sign is because of the factor of $i$). Using (\ref{3-6}) and (\ref{3-8})
\ben
{\cal N} = -\left[1-\beta_0 \cos (t-\theta)\right]
\left[1-\beta_0 \cos (t+\theta)\right]
\label{3-9}
\een
This expression immediately implies that 
\begin{itemize}

\item When $\beta_0 < 1$ one has ${\cal N} < 0$ for all $(t,\theta)$ so that the vector field $H_0(\beta_0)$ is always time-like

\item When $ \beta_0 > 1$  one has ${\cal N} < 0$ only in the region
\ben
\cos (t+\theta) < \frac{1}{\beta_0}~~~~~~~~\cos (t-\theta) < \frac{1}{\beta_0}
\label{3-10}
\een
There is a horizon described by
\bea
{\rm either} & & \cos (t+\theta) = \frac{1}{\beta_0} \nonumber \\
{\rm or } & & \cos (t-\theta) = \frac{1}{\beta_0}
\label{3-11}
\eea
When both the above conditions are satisfied, we have a bifurcation point. The horizon has a non-vanishing surface gravity and therefore characterized by a temperature.

\item When $\beta = 1$ the above horizons become {\rm extremal} since the norm now has a double zero at each of the horizons. The surface gravity and therefore the temperature vanishes.

\end{itemize}

While we have demonstrated this for $H_0 (\beta_0)$ for simplicity of presentation, the result is clearly generally valid for any $H_\mu (\beta_\mu)$. The Hamiltonian classes we discussed above therefore have a geometric meaning - time evolution by the Hamiltonian is not a time-like conformal Killing vector everywhere when $\beta \geq 1$, leading to horizons. This means that an observer whose time $\eta$ is defined by $H_\mu (\beta_\mu) = \frac{\partial}{\partial \eta}$ perceives a physical horizon in the space-time (time) $\times S^d$.

\subsection{Bulk Horizons}

For a holographic CFT the horizons in the CFT base space-time lead to horizons in the one higher dimensional bulk AdS space-times. The simplest case is for a $1+1$ dimensional CFT whose dual description would be a gravitational theory in $AdS_3$. 
To elucidate this point further, we first construct the explicit Killing vectors in global-AdS$_3$ geometry. In \cite{ddks1}, these Killing vectors are presented explicitly, which we review here. The global AdS$_3$ geometry is written as:
\begin{eqnarray} \label{ads4met}
&& ds^2 = L^2 \left( - \cosh^2\rho dt^2 + d\rho^2 + \sinh^2 \rho dY^2 \right)  \ , \\
&& Y \cdot Y \equiv Y_\mu Y^\mu = Y_1^2 + Y_2^2  = 1 \ , \quad \mu = 1,2 \ . 
\end{eqnarray} 
Let us set $L=1$. Let us write down the explicit Killing vectors corresponding to Dilatation, translation and special conformal transformation along a direction. 
\begin{eqnarray}
&& D =  \partial_t \ , \label{dil} \\
&& P_1 = i e^{- i t} \left[ Y_1 \left( \partial_\rho + \frac{1}{i} \tanh \rho \partial_t \right) - \frac{1}{\tanh\rho} \sin\theta \partial_\theta \right] \ , \label{p1} \\
&& K_1 = - i e^{i t } \left[ Y_1 \left(  \partial_\rho - \frac{1}{i} \tanh \rho \partial_t \right) - \frac{1}{\tanh\rho} \sin\theta \partial_\theta \right] \ , \label{k1}
\end{eqnarray}
where 
\begin{eqnarray}
Y_1 = \cos\theta \ , \quad Y_2 = \sin\theta \ . 
\end{eqnarray}
With the above, the following $sl(2,R)$ algebra holds:{\color{ black} \footnote{ \color{ black}Note that, here we have chosen $D \to - D$ compared to the convention in equation (\ref{2-1}), with $P$ and $K$ unchanged.}}
\begin{eqnarray}
\left[D, P_1 \right] = - i P_1 \ , \quad \left[D, K_1 \right] = i K_1  \ , \quad \left[P_1, K_1 \right] = 2 i D \ . 
\end{eqnarray}
It is also straightforward to check that the above Killing vectors satisfy the Killing equation:
\begin{eqnarray}
\nabla_A \xi_B + \nabla_B \xi_A = 0 \ .
\end{eqnarray}
Since the above two conditions are linear, any linear combination of solutions of the Killing equation is also a Killing vector. Moreover, the $sl(2,R)$ algebra does not fix the overall undetermined constant completely. To fix it completely, let us note the conformal boundary limits of (\ref{dil})-(\ref{k1}):
\begin{eqnarray}
&& \lim_{\rho \to \infty} D = \partial_t \ , \\
&& \lim_{\rho \to \infty} P_1 = \frac{1}{2} \left[ e^{-i(t-\theta)} + e^{-i(t + \theta)} \right] \partial_t - \frac{1}{2} \left[ e^{-i(t-\theta)} - e^{-i(t + \theta)}  \right] \partial_\theta \ , \\
&& \lim_{\rho \to \infty} K_1 = \frac{1}{2} \left[ e^{i(t + \theta)} + e^{ i( t - \theta)} \right] \partial_t + \frac{1}{2} \left[ e^{i( t + \theta)} - e^{ i( t - \theta)}  \right] \partial_\theta \ .
\end{eqnarray}
The above are the correct generators on the cylinder.\footnote{It can be easily checked that $L_m = z^{m+1} \partial_z$, for $m =-1, 0, 1$, on the plane maps to these generators under the plane-to-cylinder map.}

Let us now consider the general Hamiltonian:
\begin{eqnarray}
H = \alpha D + \beta P_1 + \gamma K_1 \equiv \xi^A \partial_A \ .
\end{eqnarray}
By definition, we have chosen: $D^\dagger = D$, $K^\dagger = P $ and therefore $\alpha \in {\mathbb R}$ and $\{\gamma, \beta\} \in {\mathbb C}$, with $\gamma^* = \beta$, which follows from hermiticity of the Hamiltonian.  The form of the Killing vector is given by
\begin{eqnarray}
|| \xi||^2 & \equiv & g_{AB} \xi^A \xi^B \nonumber\\
& = & -\cosh ^2(\rho ) \left(\alpha +e^{-i t} \cos (\theta ) \tanh (\rho ) \left(\beta +\gamma  e^{2 i t}\right)\right)^2-e^{-2 i t} \sin ^2(\theta ) \cosh ^2(\rho ) \left(\beta -\gamma  e^{2 i t}\right)^2 \nonumber\\
& - & e^{-2 i t} \cos ^2(\theta ) \left(\beta -\gamma  e^{2 i t}\right)^2 \ .
\end{eqnarray}
Let us define:
\begin{eqnarray}
&& Z = \beta e^{- i t} \quad \implies \quad Z^* = \beta^* e^{it} = \gamma e^{it} \ , \\
&& {\rm using} \quad \beta= \gamma^* \ . 
\end{eqnarray}
Using these, it is easy to rewrite the norm as:
\begin{eqnarray}
|| \xi||^2 &=& -\cosh ^2(\rho ) \left(\alpha +\cos (\theta ) \tanh (\rho ) \left(Z+ Z^*\right)\right)^2 -  \sin ^2(\theta ) \cosh ^2(\rho ) \left(Z - Z^*\right)^2 \nonumber\\
&& -  \cos ^2(\theta ) \left(Z - Z^*\right)^2 \label{normads3}
\end{eqnarray}
The norm is therefore a manifestly real quantity, as it should be.\footnote{Note that, from the reality condition of the norm, we can also deduce that $\alpha\in {\mathbb R}$ and $\beta = \gamma^*$. For a Hermitian Hamiltonian, the norm is guaranteed to be real.}

Let us first note that, in the limit $\rho \to \infty$ the norm in (\ref{normads3}) {\it conformally reduces}\footnote{By this, we mean that there is a conformal factor of $\cosh^2\rho$ relating $|| \xi||^2$ with {\cal N}.  Thus, the conformally compactified norm of $|| \xi||^2$ becomes identical to ${\cal N}$. This is expected since the total volume of the AdS-space diverges in the $\rho\to \infty$ limit.} to the norm in (\ref{3-9}) corresponding to the conformally Killing vector on the base space of the CFT. Before discussing the general case, it is straightforward to check some simple limiting cases. Consider $\beta = 0 = \gamma$. This yields: $|| \xi||^2 = - \alpha \cosh ^2(\rho )$ which is nowhere vanishing and therefore has no horizon. On the other hand, when $\alpha=0$, we obtain:
\begin{eqnarray}
|| \xi||^2 = -\sinh ^2(\rho ) \cos^2 (\theta )  \left(Z+ Z^*\right)^2 -  \sin ^2(\theta ) \cosh ^2(\rho ) \left(Z - Z^*\right)^2 -  \cos ^2(\theta ) \left(Z - Z^*\right)^2 \ . 
\end{eqnarray}
Now, $|| \xi||^2 = 0$ yields:
\begin{eqnarray}
\cosh(\rho) = \cos\theta \frac{\sqrt{({\rm Im}(Z))^2 + ({\rm Re}(Z))^2 }}{ \sqrt{ ({\rm Re}(Z))^2  \cos^2 \theta - ({\rm Im}(Z))^2 \sin^2\theta} } \ ,
\end{eqnarray}
as the only positive root, which is a non-extremal horizon.

\begin{figure}
\centering
\rotatebox{0}{\includegraphics*[width= 0.65 \linewidth]{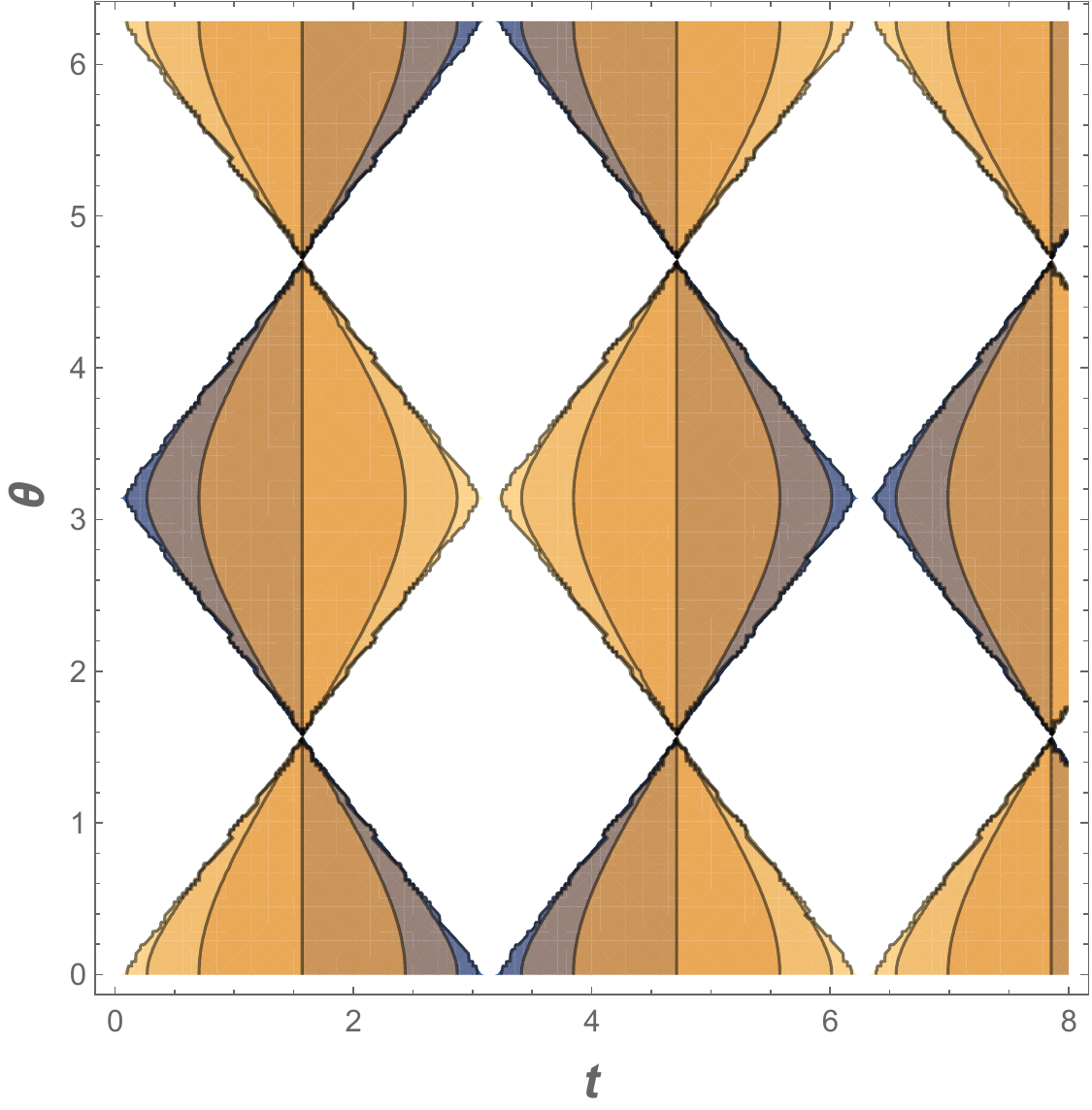}}
 \label{fig-ads3extremal} 
\caption{A contour plot showing the explicit coordinate dependence of the extremal horizon obtained in equation (\ref{exthorizon}). We have chosen here $s=0$, which corresponds to a simple shift in the $t$ coordinate. The unshaded (white) region corresponds to regions where the horizon does not exist. Evidently, these regions are bounded by null rays which yields causal diamonds in this figure. Along the $\theta$ direction, the causal diamonds are identified at $\theta=0$ and $\theta= 2\pi$. There is a periodic pattern of the causal diamonds along the Lorentzian time direction.}
\end{figure}

Let us now consider more generic cases. On the phase transition line: $\alpha^2 = 4 \beta \gamma$. This line can be parametrized by
\begin{eqnarray}
\beta = \frac{\alpha^2}{4\gamma} \ , \quad \gamma = - \frac{\alpha}{2} e^{i s} \ ,
\end{eqnarray}
where $s$ is the parameter that describes the transition line. On this line, one obtains:
\begin{eqnarray}
|| \xi||^2 = -(\alpha  \cosh (\rho ) \cos(s+t)-\alpha  \cos (\theta ) \sinh (\rho ))^2 = 0 \ , \label{exthorizon}
\end{eqnarray}
which evidently yields an extremal horizon. We have provided a representative figure in \ref{fig-ads3extremal} of an extremal horizon behaves as a function of the coordinate system. 

On the other hand, it can be checked by directly solving the equation that for a given $\alpha^2 - 4 \beta \gamma$, the solution of $|| \xi||^2=0$ depends on the sign of $4 \left( ({\rm Re}(Z))^2 + ({\rm Im}(Z))^2 \right) - \alpha^2 $. Real solutions are only possible, when $4 \left( ({\rm Re}(Z))^2 + ({\rm Im}(Z))^2 \right) - \alpha^2 > 0$. Now, it is easy to see that
\begin{eqnarray}
4 \left( ({\rm Re}(Z))^2 + ({\rm Im}(Z))^2 \right) - \alpha^2 = 4\beta \gamma - \alpha^2 \ . 
\end{eqnarray}
Therefore we obtain non-extremal solution in the heating phase, an extremal solution on the transition line and no real solution in the non-heating phase.

\section{Quaternions, Conformal Transformations and Time Evolution}
\label{ctrans}

The CFT we are interested in is defined on $S^d$. 
Since the Hamiltonians we are interested in are linear combinations of the conformal generators we can obtain the time evolution of an operator by performing the corresponding conformal transformation where the time itself is a parameter of the transformation. An efficient way to do this is to start with an operator on $R^{d+1}$, obtain the transformed operator using the form of the generators (\ref{2-3}), Weyl transform using (\ref{2-6}) and finally perform the necessary analytic continuation.

\subsection{Quaternions} 

A useful formalism to think about these processes in $d=2,3$ is the quaternionic representation of conformal transformation. In this formalism the four coordinates in $R^4$ are represented by a quaternion $Q_\mu$ 
\ben
Q_\mu = x^\mu \mathbb{I} - i \sum_{\alpha \neq \mu = 1}^3 x^\alpha \sigma_\alpha 
\label{4-1}
\een
Here $\mathbb{I}$ denotes the $2 \times 2$ identity matrix and $\sigma_\alpha$ denote the three Pauli matrices. The notation $Q_\mu$ signifies that the identity part of the quaternion is $x^\mu$. For example, 
\ben
Q_0 = \left( \begin{array}{cc}
x_0 - i x_3 & -i x_1 - x_2 \\
-ix_1+x_2 & x_0 + i x_3
\end{array}
\right)
\label{4-2}
\een
Note that we can permute the coefficients of the Pauli matrices as we wish - so that the notation $Q_\mu$ is not unique. This will change the way we extract any of the remaining $x_\mu, \nu \neq \mu$ from $Q_\mu$. One could go from one quaternionic representation to another by using the Pauli matrix algebra. For example {\color {black} 
\ben
Q_0 = x_0 \mI - ix_1 \sigma_3 - ix_2 \sigma_2 -i x_3 \sigma_1
\label{4-2-d}
\een
may be related to 
\ben
Q_1 = x_1 \mI -ix_0 \sigma_3 -ix_2\sigma_1 -ix_3 \sigma_2
\label{4-2-c}
\een
by the relation}
\ben
Q_1 = \sigma_2 Q_0 \sigma_1
\label{4-2-b}
\een

This representation is for a point on $R^4$; to obtain the quaternionic representation of a point on $R^3$ we can set $x_3=0$. In the following we will deal with $d=3$ - the results for $d=2,1$ can be obtained by setting the appropriate coordinates to zero.

In terms of quaternions the vacuum two point function of a primary field with conformal dimension $\Delta$ on $R^4$ is given by
\ben
\langle 0 | \cO(y_\mu) \cO(x_\mu) | 0 \rangle  = \frac{1}{({\rm det} [ Q_\alpha (y_\mu) - Q_\alpha(x_\mu) ])^\Delta}
\label{4-2c}
\een
The result is the same regardless of our choice of $\alpha$.

\subsection{Conformal Transformations}

A general conformal transformation on $R^4$ is a quaternionic Mobius transformation
\ben
Q_\mu \rightarrow Q_\mu^\prime = (AQ_\mu + B) (C Q_\mu + D)^{-1}
\label{4-3}
\een
where $A,B,C,D$ are themselves quaternions which obey the condition
\ben
{\rm Det}  \left( A C^{-1} D C - B C \right) = \mathbb{I}
\label{4-4}
\een
Under successive transformations
\ben
\left(\begin{array}{cc} A' & B' \\
C' & D' \end{array} \right) \cdot \left(\begin{array}{cc} A & B \\
C & D \end{array} \right) = \left(\begin{array}{cc} A'A+B'C & A'B+B'D \\
C' A+ D'C& C'B +D'D \end{array} \right)
\label{4-4a}
\een
On $R^4$ the quaternions $A,B,C,D$ are again of the form 
\ben
A = a_0 \mathbb{I} - i\sum_{i=1}^3 a_i \sigma_i,
\label{4-5}
\een 
and similarly for $B$, $C$, and $D$, where $a_0$ and $a_i$ are all real. 

We will, however, need to use this formalism to derive quantities in a $S^3$ quantization. This means we need to perform a Weyl transformation to (time) $ \times S^3$ and an analytic continuation to real time,
\bea
x^\mu = e^{it} Y^\mu
\label{2-4-1}
\eea
where $Y^\mu$ are unit vectors on $R^{d+1}$, $Y_\mu Y^\mu = 1$ describing the $S^3$.  A quaternionic Mobius transformation maintains this condition when the quantities $a_0,a_i$ (and similarly $b_0,b_i$ etc) are complex. The matrices themselves need to obey the conditions
\ben
A = D^\dagger ~~~~~~~~B = C^\dagger
\label{4-6}
\een
Under a general conformal transformation a primary operator on $R^4$ transforms as
\ben
U^\dagger \cO( x_\mu) U = [J(x)]^{\Delta/4}\cO (x^\prime_\mu)~~~~~~~J = \vert \frac{\partial x_\mu^\prime}{\partial x_\mu} \vert
\label{4-7a}
\een
where $x_\mu^\prime$ are the transformed coordinates which needs to be extracted from the $Q_\alpha^\prime$ in (\ref{4-3}).  

\subsection{Transformation by a $SU(1,1)$ subgroup}
\label{sl2r}

It was shown in \cite{ddks1} that the quaternionic transformation simplifies when the conformal transformation on $R^4$ belongs to a $SL(2,R)$ subgroup formed by $D,K_\mu,P_\mu$ for some given $\mu$ (or $SU(1,1)$ when transformed to (time) $\times S^3$). In this case, if we choose to use a quaternion $Q_\mu$ {\em with the same $\mu$} the matrices $(A,B,C,D)$ become proportional to the identity matrix $\mathbb{I}$. These can be then obtained as follows. First represent $D,K_\mu,P_\mu$ by the Pauli matrices
\bea
D \rightarrow \frac{i\sigma_3}{2}~~~~& & K \rightarrow \sigma_- ~~~~~~~P \rightarrow \sigma_+ \nonumber \\
\sigma_\pm & = & \frac{1}{2} ( \sigma_1 \pm i\, \sigma_2)
\label{4-7}
\eea
Then a general $SU(1,1)$ transformation may be represented by a $2 \times 2$ matrix of the form
\ben
V \rightarrow \left( \begin{array}{cc} a & b \\ b^\star & a^\star \end{array} \right)~~~~~~~~~~|a|^2 - |b|^2 = 1
\label{4-8}
\een
Then the transformation of the quaternion $Q_\mu$ is given by
\ben
Q_\mu \rightarrow Q_\mu^\prime = (a \mI Q_\mu -ib \mI)(ib^\star \mI Q_\mu + a^\star\mI )^{-1}
\label{4-9}
\een
To prove this, we can use the $SL(2,R)$ Baker-Campbell-Hausdorff formula to rewrite any $V$ as a product of transformations by individual $D,K_\mu$ and $P_\mu$
\ben
V \rightarrow e^{i\alpha_1 D}e^{i\alpha_2 K_\mu}e^{i\alpha_3 P_\mu}
\een
We then use the transformation of the $x_\mu$ under each of these factors and combine them. On the other hand, using the representation (\ref{4-7}) we can derive the entries $(a,b,c,d)$ Finally we compare the final result from what is obtained from (\ref{4-9}). For details, see Ref.\ \cite{ddks1}.

\subsection{Time evolution as a Conformal Transformation}

In the following, we will be interested in the conformal transformation which results from time evolution by a Hamiltonian which is some linear combination of the conformal generators
\ben
H = \sum_\alpha L_\alpha
\een
where $L_\alpha \equiv(D, K_\mu, P_\mu, M_{\mu\nu})$. The evolution operator is
\ben
 U(t) = e^{-iH t} 
\label{4-8a}
\een
The effect of this time evolution is a conformal transformation with the time $t$ being a parameter. 
The transformation acts on a quaternion according to (\ref{4-3}) where the quaternionic parameters $A,B,C,D$ are functions of time - we will therefore denote these as $A(t), B(t), C(t), D(t)$. On a primary operator this action is given by (\ref{4-7}) with $U \rightarrow U(t)$

Consider the two point function of the scalar primary operator $\cO$ under such a time evolution. At time $t=0$ these two points are given, in terms of coordinates on $R^4$
\bea
\vx_1 & \equiv & (1,0,0,0) \nonumber \\
\vx_2 & \equiv & (\cos \theta_2, \sin \theta_2 \cos \phi_2, \sin \theta_2 \sin \phi_2 \cos \Psi_2, \sin \theta_2 \sin \phi_2 \sin \Psi_2)
\label{4-9a}
\eea
The unequal time correlator with the second operator placed at time $t$ is then given by
\bea
\Gamma (t) & = & \langle 0 | U^\dagger (t) \cO(\vx_2) U(t) \cO(\vx_1) |0\rangle \nonumber \\
& = & \frac{J(\vx_2)^{\Delta/4}}{({\rm det} [Q_\alpha^\prime(\vx_2) - Q_\alpha(\vx_1)])^\Delta}
\label{4-9b}
\eea
where the jacobian is given by 
\ben
J(\vx_2) = ({\rm det} [C (t) Q_\alpha (\vx_2) +D (t)])^{-4}
\label{4-10}
\een
Note that the point $\vx_1$ is not transformed. 
The time dependence of this quantity is determined entirely by the time dependence of the quaternionic parameters $A(t), B(t), C(t), D(t)$. 

We are finally interested in correlators on (time) $\times S^3$. This is obtained by performing a Weyl transformation on (\ref{4-9b}), and an analytic continuation $\tau = it$ to obtain
\ben
C(t, \theta_2,\phi_2,\psi_2)) = e^{it \Delta} \Gamma (t)
\label{4-11}
\een
Note that the CFT vacuum is annihilated by the Hamiltonians $H_\mu (\beta_\mu)$. However the spectrum is different from the standard Hamiltonian $H_0 = 2iD$, resulting in a time dependence of the unequal time correlator. Similarly, the energy density and the fidelity of primary states also depend non-trivially on time.

\subsection{Time evolution by a single $H_\mu$}
\label{single}

When the time evolution is by a single Hamiltonian $H_\mu (\beta_\mu)$ we can use the simplification discussed in sub-section (\ref{sl2r}). Consider for example time evolution by $H_0 (\beta_0)$. It is then useful to choose a quaternionic representation $Q_0 (x_\mu)$, on which the time evolution acts via (\ref{4-9}). The result for the matrix $V$ in (\ref{4-8}) can be easily obtained
\ben
V \rightarrow \left( \begin{array}{cc} a(t) & b(t) \\
b^\star (t) & a^\star (t) \end{array}
\right)
\label{5-2}
\een
where
\bea
a(t) =  \cos (\nu_0 t) + \frac{i}{\nu_0}\sin (\nu_0 t) & & b(t) = \frac{\beta_0}{\nu_0} \sin (\nu_0 t) \nonumber \\
\nu_0 & = & \sqrt{1-\beta_0^2}
\label{5-3}
\eea
The two point function can be now readily obtained. Using (\ref{4-9b})-(\ref{4-11}) we obtain
\bea
\frac{C(t)}{C(0)} & = & \frac{[2(1-\cos\theta_2)]^\Delta}{\left[ \left( a^2 + (a^\star)^2 - b^2 - (b^\star)^2  - 2i (ab^\star - ba^\star) \right) - 2\cos\theta_2 \left( |a|^2 + |b|^2 + i (ab-a^\star b^\star) \right)\right]^\Delta} \nonumber \\
& = & \frac{[1-\cos\theta_2]^\Delta}{\left[ \cos^2(\nu_0 t) (1-\cos \theta_2) - \frac{\sin^2(\nu_0t)}{\nu_0^2} \{ (1+ \beta_0)^2 + (1-\beta_0)^2 \cos \theta_2\} \right]^\Delta}
\label{5-4}
\eea
The time dependence of corrrelators of the operator $\cO$ is clearly contained entirely in the functions $a(t)$ and $b(t)$. The expressions (\ref{5-4}) immediately reveals that the nature of the time evolution depends on whether $\beta_0 \lesseqgtr 1$. 

\begin{itemize}

\item{} For $\beta_0 < 1$ these are oscillatory functions of time, so that e.g. unequal time corrrelators of operators $\cO$ will oscillate in time for ever.  The time period is given by
\ben
T_0 = \frac{2\pi}{\sqrt{1-\beta_0^2}}
\een
This is a non-heating phase.

\item{} For $\beta_0 = 1$ we have $a(t) = 1+it, b(t) =t$: this leads to power law decay of correlators. This is a critical phase

\item{} For $\beta_0 > 1$ the functions $a(t)$ and $b(t)$ are hyperbolic functions. This leads to
\ben
\frac{C(t)}{C(0)} \sim e^{-2\Delta t \sqrt{\beta_0^2-1} }~\frac{[2(\beta_0^2-1)\cos\theta_2]^\Delta}{[1+ \beta_0 (1+\cos\theta_2)]^\Delta}
\een
The correlator therefore decays exponentially in time, characteristic of a ``heating'' phase. Note, however, the inhomogeneity persists at arbitrarily late times. 

\end{itemize}
 
The three phases described above are {\em equilibrium} phases: they are properties of time evolution due to a {\em single} Hamiltonian. As shown in \cite{ddks1} the fidelity of a primary state, as well as expectation values of the energy density as well as primary operators in such a state exhibit oscillations, {\color{ black} power law} decays or exponential decays for $\beta \lesseqgtr 1$.

\section{ Floquet Dynamics}
\label{fldyn} 

In this section, we shall discuss several Floquet drive protocols. These can be divided into two categories. The first 
class of drives involves use of a single $SU(1,1)$ group and is discussed in Sec.\ \ref{sufl} while the second class involves 
the full $SO(4,2)$ group and is discussed in Sec.\ \ref{fullfl}.

\subsection{$SU(1,1)$ drives}  
\label{sufl}

The simplest Floquet dynamics is described by a two step process described by the Hamiltonian (\ref{1-4}).
As discussed earlier the effect of time evolution on quaternions becomes simple if we use the quaternion $Q_0$. At the end of a single cycle from $t=0$ to $t=T=T_1+T_2$ is described by a matrix $V$ defined in (\ref{4-8})
\ben
V = \left( \begin{array}{cc} a(T_1,T_2) & b(T_1,T_2) \\
b^\star (T_1,T_2) & a^\star (T_1,T_2) \end{array}
\right)
\label{6-1}
\een
where the matrix elements are proportional to the $2 \times 2$ identity matrix $\mathbb{I}$
\bea
a(T_1,T_2) = [ \cos (\nu_0 T_2) + \frac{i}{\nu_0}\sin (\nu_0 T_2) ] e^{iT_1} \mathbb{I}& & b(T_1,T_2) = e^{-iT_1}\frac{\beta_0}{\nu_0} \sin (\nu_0 T_2) \mathbb{I}\nonumber \\
\nu_0 & = & \sqrt{1-\beta_0^2}
\label{6-2}
\eea
Let us denote
\bea
a_r  & = & {\rm Re} [a (T_1,T_2)]~~~~~~~b_r = {\rm Re} [b (T_1,T_2)] \nonumber \\
a_I  & = & {\rm Im} [a (T_1,T_2)]~~~~~~~b_I  = {\rm Im} [b (T_1,T_2)]
\label{6-3}
\eea
Expanding the matrix (\ref{6-1}) in terms of Pauli matrices it is now straightforward to obtain the resulting transformation after $n$ steps
\ben
V_n = \left( \begin{array}{cc} a_n(T_1,T_2) & b_n(T_1,T_2) \\
b_n^\star (T_1,T_2) & a_n^\star (T_1,T_2) \end{array}
\right)
\label{6-4}
\een
Expressions for the physical quantities at the end of $n$ cycles is clearly determined by these functions $a_n,b_n$. For example, the correlation function at times $nT$ are given by (\ref{5-4}) with the replacement $a \rightarrow a_n$ and $b \rightarrow b_n$. 

The coefficients $a_n(T_1,T_2)$ and $b_n(T_1,T_2)$ depend on $a,b$ only for the $SU(1,1)$ algebra. Therefore one can use the Pauli matrix representation to calculate them directly.
The nature of this stroboscopic response is determined by the value of $a_r (T_1,T_2,\beta_0)$

\begin{itemize}

\item{} For $a_r^2 < 1$,
\bea
a_n (T_1,T_2) & = & \cos (n~\arccos(a_r) )+ i \frac{a_I}{\sin (\arccos(a_r))} \sin (n~\arccos(a_r)) \nonumber \\
b_n & = & \frac{b_r (T_1,T_2)+ i b_I (T_1,T_2)}{\sin (\arccos(a_r))} \sin (n~\arccos(a_r)).
\label{6-5}
\eea
 This response is oscillatory. Note that the time period is a complicated function of the parameters
\ben
T_p = \frac{2\pi}{\arccos(a_r)}
\label{6-10a}
\een
This is a "non-heating" phase.

\item{} 
For $a_r^2=1$ we have
\ben
a_n (T_1,T_2) =1 + i n a_I ~~~~~~~~~
b_n (T_1,T_2) = n[b_r + ib_I]
\label{6-7}
\een
This means that the response now decays {\color{ black} as a power law} in the number of cycles $n$. This is a critical phase.

\item{}
Finally for $a_r^2 > 1$ we have
\bea
a_n & = &  \cosh (n~ {\rm {arccosh}} (a_r)) + i \frac{a_I}{\sinh ({\rm {arccosh}}(a_r ))} \sinh (n ~ {\rm {arccosh}} (a_r)) \nonumber \\
b_n & = & \frac{b_r + i b_I}{\sinh ({\rm {arccosh}}(a_r))} \sinh ( n ~{\rm {arccosh}}(a_r))
\label{6-8}
\eea
The response decays exponentially for large $n$. This is a ``heating" phase.
\end{itemize}

An easier and perhaps more useful way to derive the nature of the $n$ dependence of $(a_n,b_n)$ is to go to a diagonal representation of the matrix $V$ in (\ref{6-1}). It is easily seen that the eigenvalues are $\lambda_\pm$ where 
\ben
\lambda_\pm = a_r \pm \sqrt{a_r^2 - 1}
\label{6-10}
\een
Thus
\bea
a_r & < & 1~~~~~~~~~~\lambda_\pm = \exp [{\pm~ i~ {\rm arccos} (a_r)}] \nonumber \\
a_r & = & 1~~~~~~~~~~\lambda_\pm = 1 \nonumber \\
a_r & > & 1~~~~~~~~~~\lambda_\pm = \exp [ {\pm ~ {\rm arccosh} (a_r)}]
\eea
If $P$ denotes the matrix which diagonalizes $V$, i.e. 
\ben
\Lambda \equiv \left( \begin{array}{cc} \lambda_+ & 0 \\
0 & \lambda_- \end{array} \right) = P^{-1} V P
\een
we immediately get
\ben
V_n = P \Lambda^n P^{-1} = P \left( \begin{array}{cc} \lambda_+^n & 0 \\
0 & \lambda_-^n \end{array} \right) P^{-1} \label{eq:Vn}
\een
so that the $n$ dependence in $V_n$ is entirely contained in $\lambda_\pm^n$. This immediately shows that when $a_r > 1$ the behavior is exponential and when $a_r < 1$ the behavior is oscillatory. $a_r =1$ is subtle and needs to be considered as a limit, which leads to (\ref{6-7}).

These three phases are {\em dynamical} phases, since they are determined by the parameters in the Hamiltonian at each step, as well as the frequency of the drive. The system can be made to transition from a heating to a non-heating phase by changing $T_1,T_2$ {\em regardless of the value of $\beta_0$}.

The Floquet Hamiltonian in each of this phases is a linear combination of the $D,K_0,P_0$ which form a $SU(1,1)$ subalgebra. This has been explicitly obtained in \cite{ddks1}. In fact the dynamical phase is the equilibrium phase of the Floquet Hamiltonian which of course depends on $T_1,T_2$. Significantly, the dynamical phase is determined by a single number, $a_r(T_1,T_2,\beta_0)$. This quantity which determines the type of conjugacy class of the $SU(1,1)$ transformation, with $a_r^2 < 1$ being the elliptic class, $a_r^2=1$ being the parabolic class and $a_r^2 > 1$ being the hyperbolic class. The phase structure is identical to $SU(1,1)$ drives in $1+1$ dimensional conformal field theories, which has been extensively discussed in the literature.

\subsection{More general $SO(4,2)$ Floquet drives}
\label{fullfl}

Multi-step drives with different Hamiltonians $H_\mu (\beta_\mu)$ in different steps (as in (\ref{1-5})) would result in Floquet Hamiltonians which are linear combinations of generators of the conformal group. {\color{ black} However unlike the 2D case there is no one-to-one correspondence of the types of conjugacy classes of the Euclidean conformal group with the stroboscopic response of the system. For conformal transformations on the Euclidean plane $R^4$, these conjugacy classes are known in terms of the elements of the equivalent quaternionic Mobius transformation $SL(2,H)$ \cite{parker-short}. Since we need the transformations of the coordinates on (time) $\times S^d$, the matrices $A,B,C,D$ will have complex entries, as is already evident in the case of $SU(1,1)$ drives in the previous section. For the complex case which arises naturally in the Lorentzian setting, the correspondence with conjugacy classes of $SL(2,H)$ does not hold.}
In this sub-section we shall consider drives that belong to a $SO(2,2)$ subalgebra of the full conformal group for both $2$-step and $3$-step drive protocols. 

\subsubsection{A 2 step cycle}
\label{2-step}

Let us first consider a protocol given by (\ref{1-5}). The Floquet Hamiltonian will be a linear combination of $D,K_1,P_1,K_0,P_0,M_{01}$ which forms a $SO(2,2)$ subalgebra.

We now need to use two different quaternionic representation of the coordinates on $R^4$, $Q_0,Q_1$ which are defined in (\ref{4-2-d}) and (\ref{4-2-c}). As we discussed, in a step where the Hamiltonian is $H_\mu(\beta_\mu)$ for some given $\mu$, the quaternionic entries of the transformation matrix are proportional to the identity matrix $\mathbb{I}_{2 \times 2}$. Therefore for the protocol (\ref{1-5}) it is convenient to use the representation $Q_1$. Starting at $t=0$ the quaternion at time $t=T_1$ is then given by
\ben
Q_1^\prime = (a_1 Q_1 -ib_1)(ib_1^\star Q_1 + a_1^\star)^{-1}
\label{7-1}
\een
where the complex numbers $a_1,b_1$ are given by (following (\ref{5-3}))
\bea
a_1 = \cos(\nu_1 T_1)  + \frac{i}{\nu_1} \sin (\nu_1 T_1)~~& & ~~~~b_1 = \frac{\beta_1}{\nu_1} \sin (\nu_1 T_1) \nonumber \\
\nu_1 & = & \sqrt{1-\beta_1^2}
\label{7-2}
\eea
This means that in the quaternionic representation $Q_0$ we have, using $Q_1=\sigma_2 Q_0 \sigma_1$
\ben
Q_1^\prime = (a_1\sigma_2 Q_0 -ib_1\sigma_1)(ib_1^\star \sigma_2 Q_0 +a_1^\star \sigma_1)^{-1}
\label{7-3}
\een
Using (\ref{4-2-b}) once more we get the quaternion $Q_0^\prime$ at $t = T_1$ 
\ben
Q_0^\prime = (a_1 Q_0 -b_1\sigma_3)(-b_1^\star \sigma_3 Q_0 +a_1^\star)^{-1}
\label{7-4}
\een
In the next step of the cycle, from $t=T_1$ to $t=T_1+T_2$ we use the fact that since the Hamiltonian is now $H_0 (\beta_0)$ the quaternion $Q_0^{\prime\prime}$ at the end of the cycle at $t=T_1+T_2$ is given by 
\ben
Q_0^{\prime\prime} =  (a_2 Q_0^\prime -ib_2)(ib_2^\star Q_0^\prime + a_2^\prime)^{-1} 
\label{7-5}
\een
where
\bea
a_2 = \cos(\nu_0 T_2)  + \frac{i}{\nu_0} \sin (\nu_0 T_2)~~& & ~~~~b_2 = \frac{\beta_0}{\nu_0} \sin (\nu_0 T_2) \nonumber \\
\nu_0 & = & \sqrt{1-\beta_0^2}
\label{7-6}
\eea
Note that
\ben
|a_1|^2 - |b_1|^2 = |a_2|^2 - |b_2|^2 = 1
\label{7-6-a}
\een
Combining (\ref{6-5}) and (\ref{6-4}) we get the quaternion at the end of the cycle by
\ben
Q_0^{\prime\prime} = Q_0^{(1)} = \left[ (f_1 + ig_1 \sigma_3)Q_0 + (f_2 -ig_2 \sigma_3) \right]\left[ (f_2^\star + i g_2^\star \sigma_3)Q_0 + (f_1^\star - ig_1^\star \sigma_3) \right]^{-1}
\label{7-7}
\een
where
\bea
f_1 & = & a_1 a_2~~~~~~~~~f_2 = -i b_2 a_1^\star \nonumber \\
g_1 & = & b_1^\star b_2~~~~~~~~~g_2 = -ia_2b_1
\label{7-8}
\eea
Note that these are (complex) numbers, not matrices.

\begin{figure}
\centering
\rotatebox{0}{\includegraphics*[width= 0.45 \linewidth]{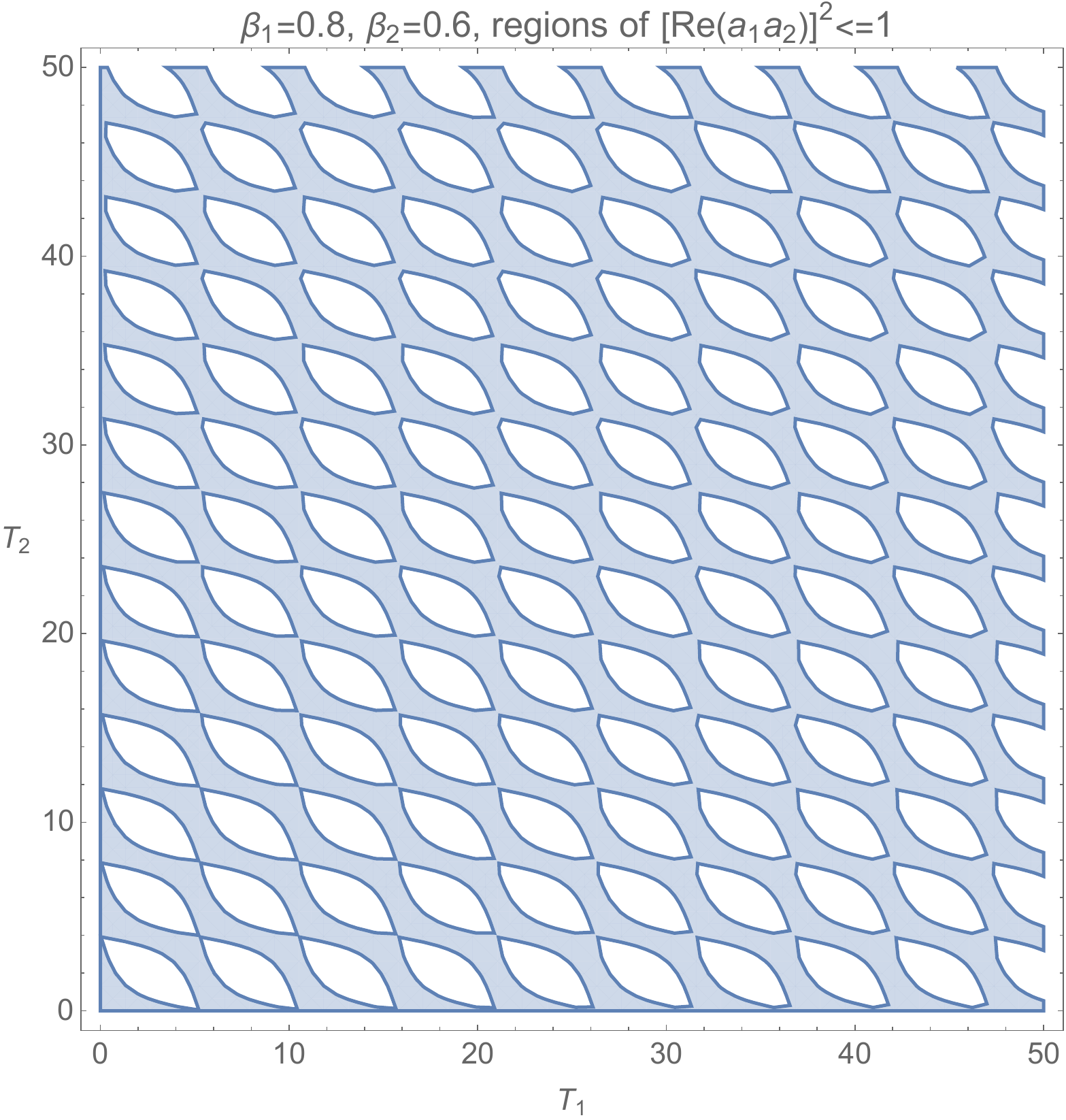}}
\rotatebox{0}{\includegraphics*[width= 0.45 \linewidth]{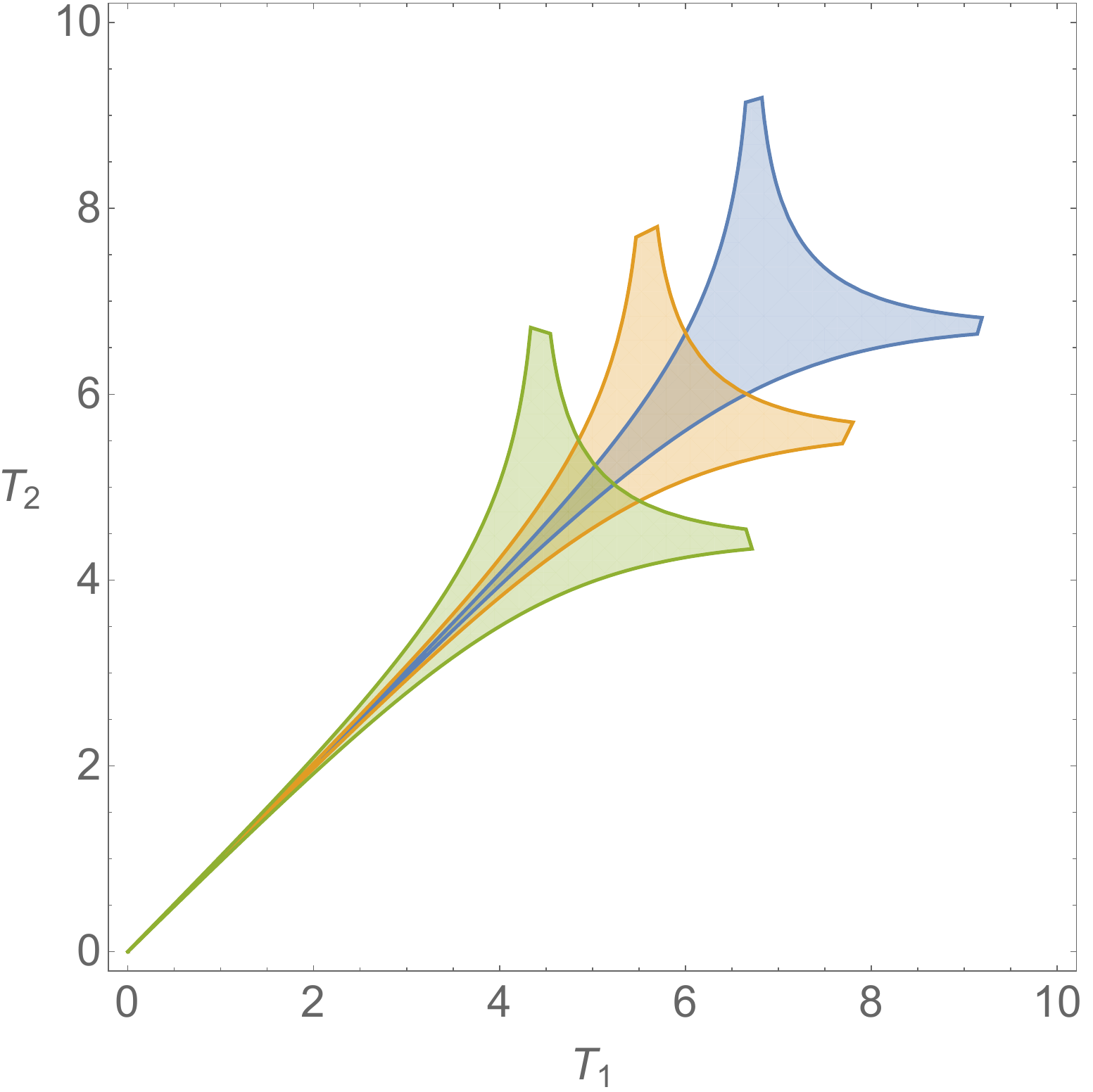}}
\caption{Left : Pure phase eigenvalue plots for parameters $\beta_1 = 0.8, \beta_0 = 0.6$ are shown in the blue region. Right : Blue region indicates pure phase eigenvalues for $\beta_0 = \beta_1 = \sqrt{2} - 10^{-6}$, orange for $\beta_0 = \beta_1 = \sqrt{2} - 10^{-5}$ and green when $\beta_0 = \beta_1 = \sqrt{2} - 10^{-4}$.  \label{fig-bidirec-1} }
\end{figure}

To explore the stroboscopic properties of this protocol we need to iterate this procedure $n$ times. This is easier to accomplish if we first diagonalize the matrix
\ben
\left( \begin{array}{cc}
f_1 + ig_1 \sigma_3 & f_2 - ig_2 \sigma_3 \nonumber \\
f_2^\star +i g_2^\star \sigma_3 & f_1^\star - ig_1^\star \sigma_3
\end{array}\right)
\label{7-9}
\een
is diagonal. The eignevalues of this matrix are $\lambda_\pm$ and $\mu_\pm$ where
\bea
\lambda_\pm & = & {\rm Re} (f_1+ig_1) \pm \sqrt{[{\rm Re} (f_1+ig_1)]^2 - (|f_1 + ig_1|^2 - |f_2 -ig_2|^2)} \nonumber \\
\mu_\pm & = & {\rm Re} (f_1-ig_1) \pm \sqrt{[{\rm Re} (f_1-ig_1)]^2- (|f_1 -ig_1|^2 - |f_2 +ig_2|^2)}
\label{7-10}
\eea
It follows from (\ref{7-8}) and (\ref{7-6-a}) that 
\bea
 |f_1 + ig_1|^2 - |f_2 -ig_2|^2  =   |f_1 - ig_1|^2 - |f_2 + ig_2|^2  = (|a_1|^2 - |b_1|^2)(|a_2|^2 - |b_2|^2) = 1
\label{7-11}
\eea
Furthermore $b_1$ and $b_2$ are real (from (\ref{7-2}) and (\ref{7-6})) so that 
\ben
{\rm Re} (f_1 + ig_1) = {\rm Re} (f_1 - ig_1) = {\rm Re} (a_1a_2)
\label{7-12}
\een
This means that out of the four eigenvalues two of them coincide,
\ben
\lambda_\pm = \mu_\pm = {\rm Re} (a_1a_2) \pm \sqrt{({\rm Re} (a_1a_2))^2 -1}
\label{7-13}
\een
As before, the stroboscopic behavior is determined by a single quantity, $ {\rm Re} [a_1 (T_1,\beta_1) a_2 (T_2,\beta_0)]$ which is now a function of $T_1, T_2, \beta_1, \beta_0$. 
\begin{itemize}

\item{} When $ {\rm Re} [a_1 a_2] < 1$ \footnote{We are not writing the variables on which the $a_i$ depend to avoid clutter} the response is an oscillatory function of the number of cycles $n$ with a period given by
\ben
T_p = \frac{2\pi}{{{\rm arccos} ({\rm Re}~a_1a_2)}}
\een

\item{} When $ {\rm Re} [a_1  a_2 ] = 1$ the response is a {\color{ black} polynomial} in $n$.

\item{} When $ {\rm Re} [a_1 a_2] > 1 $ the response decays exponentially in $n$.

\end{itemize}
In Fig.\ \ref{fig-bidirec-1} we show the regions in the $T_1, T_2$ plane wherein we can have elliptic phases in steady state.
The phase boundary in the four parameter space is given by the hypersurface:
\begin{align}
\cos \left( T_1 \sqrt{1 - \beta_1^2} \right) \cos \left( T_2 \sqrt{1- \beta_0^2}\right) - \frac{ \sin \left( T_1 \sqrt{1 - \beta_1^2} \right)\sin \left( T_2 \sqrt{1- \beta_0^2}\right)  }{\sqrt{1-\beta_1^2} \sqrt{1-\beta_0^2} }  &=1. \label{eq:2stepbndy}
\end{align}

\subsubsection{A 3 step cycle}
\label{3-step-1}

We now consider a 3 step cycle given by
{\color{ black}
\bea
H  & = & 2iD~~~~~~~~~~~~~~nT \leq t \leq nT+T_0 \nonumber \\
H  &  = & H_1 (\beta_1)~~~~~~~nT + T_0 \leq t \leq nT + T_0+ T_1 \nonumber \\
H  &  = & H_0 (\beta_0)~~~~~~~nT + T_0+ T_1 \leq t \leq nT + T_0+ T_1 + T_2 = (n+1)T
\label{8-1}
\eea
}
where the time period is $T = T_0+T_1+T_2$. Following the discussion in the previous section, we choose the quaternionic representation $Q_1$ for the first two steps of a single cycle, convert the result to a $Q_0$ representation and use this representation for the last step. At the end of the first two steps, the quaternion $Q_1$ is given by
\ben
Q_1^\prime = (c_1 Q_1 -id_1)(id_1^\star Q_1 + c_1^\star)^{-1}
\label{8-2}
\een
where (following (\ref{6-2})),
\bea
c_1(T_0,T_1) = [ \cos (\nu_1 T_1) + \frac{i}{\nu_1}\sin (\nu_1 T_1) ] e^{iT_0} \mathbb{I}& & d_1(T_0,T_1) = e^{-iT_0}\frac{\beta_1}{\nu_1} \sin (\nu_1 T_1) \mathbb{I}\nonumber \\
\nu_1 & = & \sqrt{1-\beta_1^2}
\label{8-3}
\eea
The result at the end of the third step of the cycle can be obtained exactly as in (\ref{2-step})
\ben
Q_0^{\prime\prime} = Q_0^{(1)} = \left[ (p_1 + iq_1 \sigma_3)Q_0 + (p_2 -iq_2 \sigma_3) \right]\left[ (p_2^\star + iq_2^\star \sigma_3)Q_0 + (p_1^\star - iq_1^\star \sigma_3) \right]^{-1}
\label{8-4}
\een
where
\bea
p_1 & = & c_1 c_2~~~~~~~~~p_2 = -i d_2 c_1^\star \nonumber \\
q_1 & = & d_1^\star d_2~~~~~~~~~q_2 = -ic_2d_1
\label{8-5}
\eea
and
\bea
c_2 = \cos(\nu_0 T_2)  + \frac{i}{\nu_0} \sin (\nu_0 T_2)~~& & ~~~~d_2 = \frac{\beta_0}{\nu_0} \sin (\nu_0 T_2) \nonumber \\
\nu_0 & = & \sqrt{1-\beta_0^2}
\label{8-6}
\eea
As above we also have
\bea
 |p_1 + iq_1|^2 - |p_2 -iq_2|^2  =   |p_1 - iq_1|^2 - |p_2 + iq_2|^2  = (|c_1|^2 - |d_1|^2)(|c_2|^2 - |d_2|^2) = 1
\label{8-7a}
\eea
The eigenvalues of the matrix which represents the transformation at the end of a single cycle are therefore
\bea
\lambda_\pm & = & {\rm Re} (p_1+iq_1) \pm \sqrt{[{\rm Re} (p_1+iq_1)]^2 - 1} \nonumber \\
\mu_\pm & = & {\rm Re} (p_1-iq_1) \pm \sqrt{[{\rm Re} (p_1-iq_1)]^2- 1}
\label{8-7}
\eea
Unlike the results of the previous section, $d_1,d_2$ are not real  so that
${\rm Re} (p_1+iq_1) \neq {\rm Re} (p_1-iq_1)$. Therefore there are two distinct conditions which determine the nature of these eigenvalues. Consequently, there are a multitude of phases

\begin{enumerate} 

\item{} The response is purely exponential when both  $[{\rm Re} (p_1-iq_1)]^2 > 1$ and $[{\rm Re} (p_1+iq_1)]^2 > 1$.

\item{} A purely oscillatory phase when both  $[{\rm Re} (p_1-iq_1)]^2 < 1$ and $[{\rm Re} (p_1+iq_1)]^2 < 1$.

\item{} A parabolic phase when both  $[{\rm Re} (p_1-iq_1)]^2 = 1$ and $[{\rm Re} (p_1+iq_1)]^2 = 1$

\item{} A phase when one of the quantities  $[{\rm Re} (p_1-iq_1)]^2$ and $[{\rm Re} (p_1+iq_1)]^2 $ is greater than unity while the other is less than unity. {\color{ black}In this case now the $4 \times 4$ matrix $V_n$ \eqref{eq:Vn} has 4 entries that oscillate with $n$, while 4 other exponentially grow as function of $n$. See App. \ref{app:Vn} for an explicit form of $V_n$. The two-time correlator for this case exhibits an exponential decay with stroboscopic time $n$ with oscillations on top of it, see Fig. \ref{fighybrid} for an illustrative example.} \label{item4}

\end{enumerate}

\begin{figure}
\centering
\rotatebox{0}{\includegraphics*[width= 0.45 \linewidth]{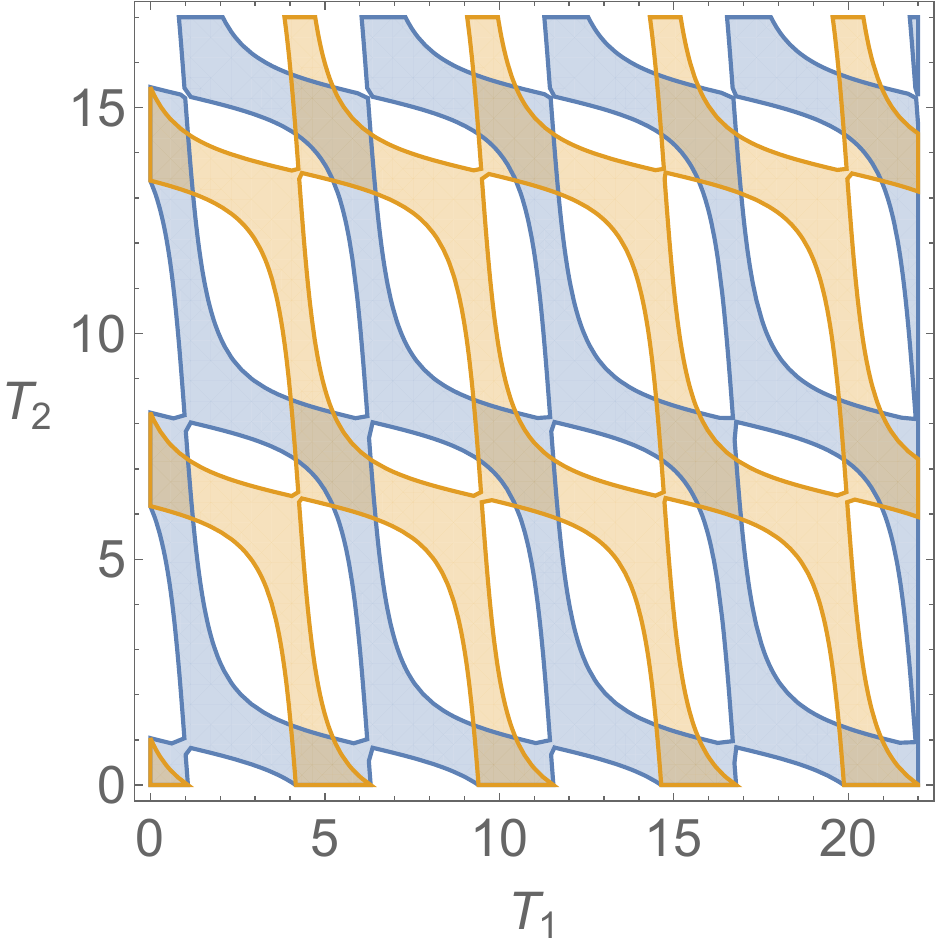}}
\rotatebox{0}{\includegraphics*[width= 0.45 \linewidth]{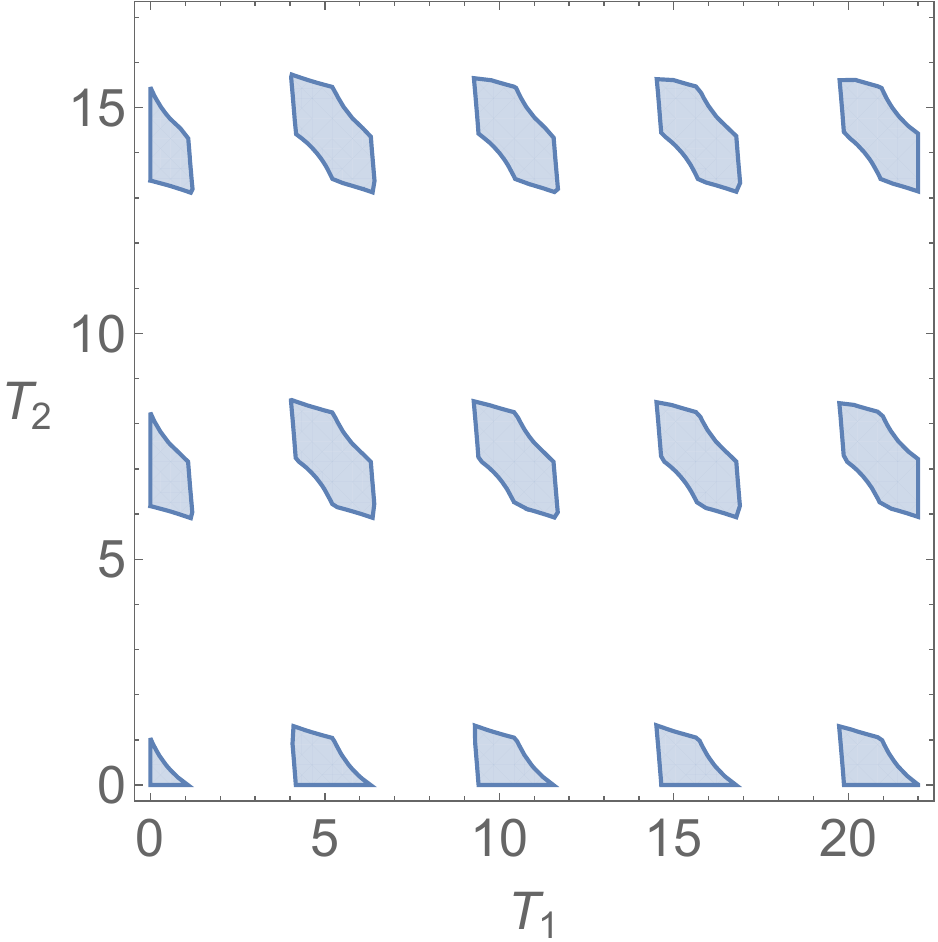}}
\caption{Eigenvalue plots for parameters $\beta_0 = 0.9, \beta_1 = 0.8, T_0 = 11$. Left : Blue regions indicate $\lambda_\pm$ are pure phase, whereas Orange indicates $\mu_\pm$ are pure phase. Right : The intersection region where all 4 eigenvalues are pure phase. These are the regions in the parameter space which will result in {\em elliptic} steady state. \label{fig3} }
\end{figure}
There are more conditions to be met for pure phase eigenvalues, and even though the number of parameters have increased, the allowed regions have shrunk in the three step bidirectional drive as compared with the two step case. This feature, shown in Fig.\ \ref{fig3}, continues to hold as a generic feature, such that the elliptic phase becomes rare.

\begin{figure}
	\centering
	\includegraphics*[width= 0.75 \linewidth]{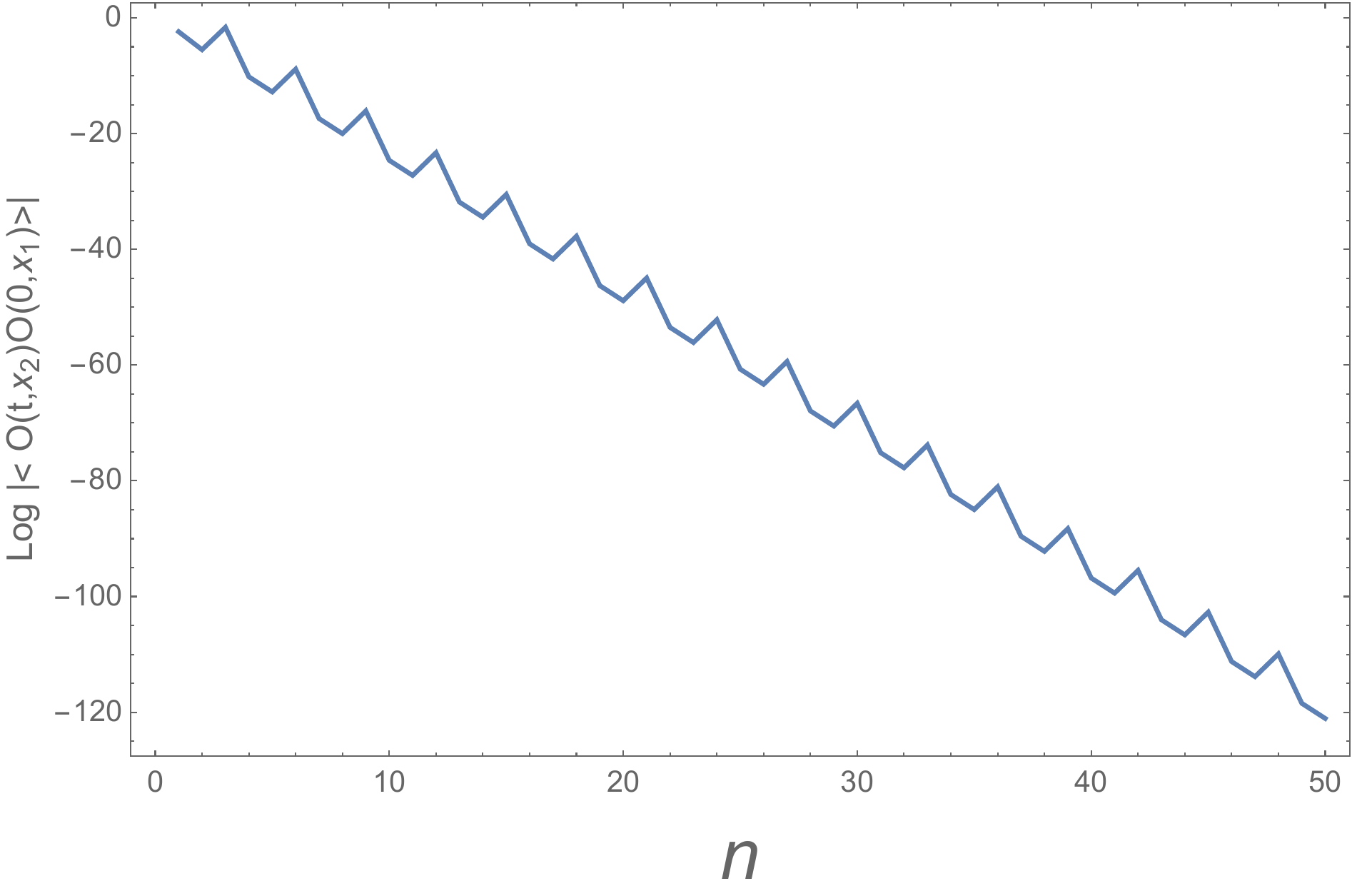}
	\caption{{\color{ black}Correlation function in the hybrid phase. The parameters take values, $p_1 = \tfrac{1}{2} + 2\,i , \, q_1 = -i, \, p_2 = 1 + \sqrt{3}\, i, \, q_2 = \tfrac{i}{2}$.} \label{fighybrid} }
\end{figure}

\subsection{A general three directional Floquet drive}
\label{three-step}
In this subsection we generalize the drives of  \S \ref{2-step} and \S \ref{3-step-1} to a three directional periodic deformation, \ie the following protocol: 
\begin{align}
H_1 &= 2i D + i \beta_1 ( K_i + P_i ) , \,\, 0 \leq t \leq T_1, \\
H_2 &= 2i D + i \beta_2 ( K_k + P_k ) , \,\, T_1 \leq t \leq T_1+T_2, \\ 
H_3 &= 2i D + i \beta_2 ( K_p + P_p ) , \,\, T_1 +T_2 \leq t \leq T_1+T_2+T_3.
\end{align}
Here the deformations are first in the $i$, then in the $k$ and finally in the $p$-th direction, which are distinct. For each direction we have complex numbers similar to \eqref{7-2} that we denote by $a_i, b_i$ for $i=1,2,3$. 
Next we define the indices $r$ and $s$ such that
$
- i \, \sigma_s \, \sigma_p \, \sigma_r = 1.
$
Notice that $p \neq k$  implies that $k$ is either equal to $r$ or $s$. Let us assume in what follows below that $k=r$. Next, via a series of algebraic steps similar to the simpler drives, the final quaternionic transformation can be expressed as: 
\begin{align}
Q_p'' &= \bigg[ A \, Q_p - i\, B \bigg]\cdot \bigg[ i \, C \, Q_p + D\bigg]^{-1},
\end{align}
where we find:
\begin{align}
A &= {a}_1 \, {a}_2 \, {a}_3 \,\mathbb{I} - i \, {b}_1^* \, {b}_2 \,{a}_3\, \sigma_k + i \, {a}_1\, {b}_2^* \, {b}_3 \, \sigma_p + i \, {b}_1^* \, {a}_2^* \, {b}_3\, \sigma_s,  \\
B&= {a}_1^* \, {a}_2^* \,{b}_3\,  \mathbb{I} -i\, {b}_1 \, {a}_2^* \, {b}_3 \, \sigma_k  - i \, {a}_1^*\, {b}_2 \, {a}_3 \, \sigma_p -  i \, {b}_1 \, {a}_2 \,{a}_3\, \sigma_s,   \\
C&={a}_1 \, a_2 \, {b}_3^* \,\mathbb{I} - i \, {b}_1^* \, {b}_2 \,{b}_3^*\, \sigma_k + i \, {a}_1 \, {b}_2^* \, {a}_3^* \, \sigma_p + i \, {b}_1^* \, {a}_2^* \, {a}_3^*\, \sigma_s,\\
D &={a}_1^* \, {a}_2^* \,{a}_3^*\,  \mathbb{I} -i\, {b}_1\, {a}_2^* \, {a}_3^* \, \sigma_k  - i \, {a}_1^*\, {b}_2 \, {b}_3^* \, \sigma_p -  i \, {b}_1 \, {a}_2 \,{b}_3^*\, \sigma_s. \label{set2}
\end{align}

\begin{figure}[htbp!]
\centering
\rotatebox{0}{\includegraphics*[width= 0.46 \linewidth]{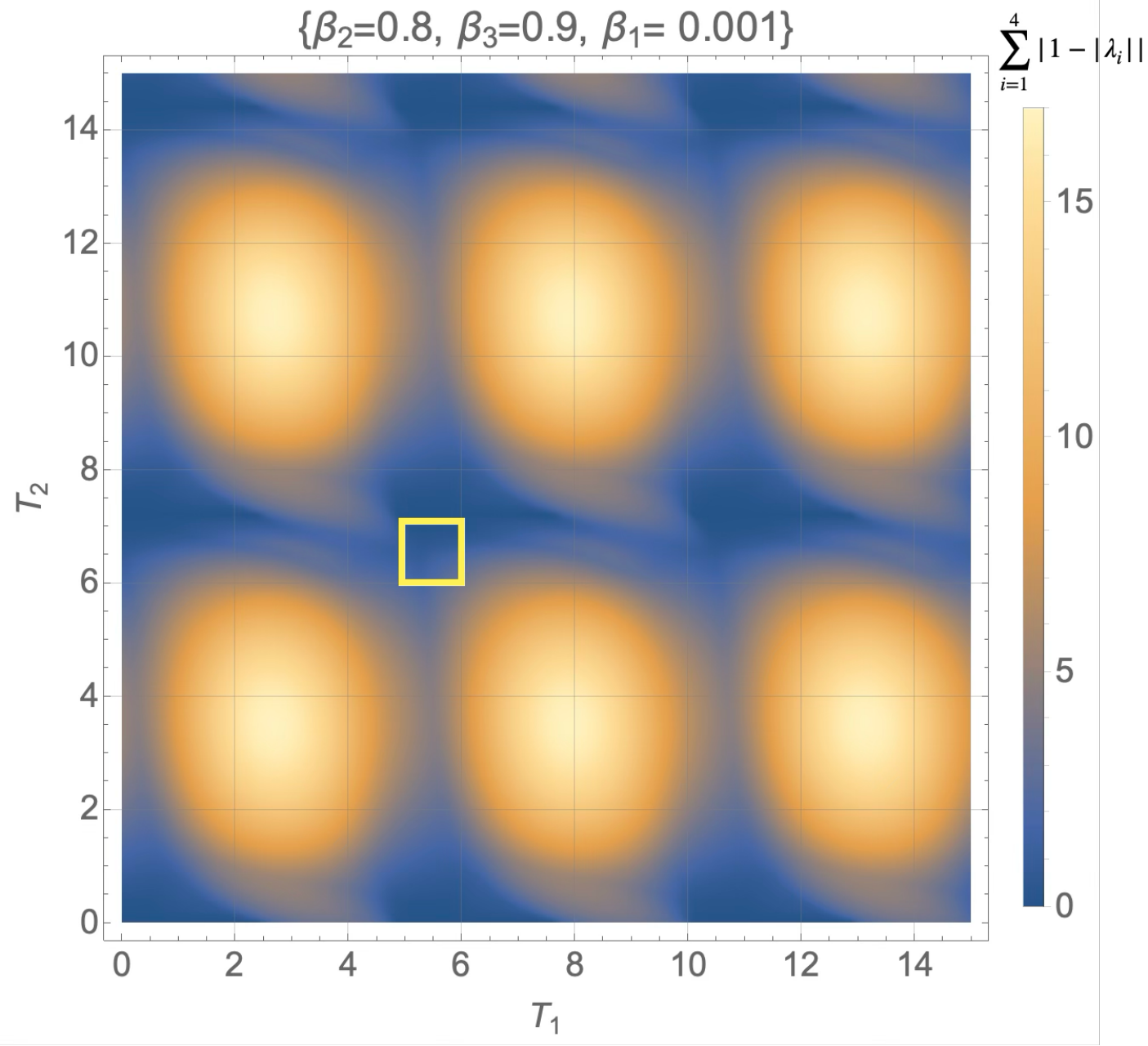}}
\rotatebox{0}{\includegraphics*[width= 0.48 \linewidth]{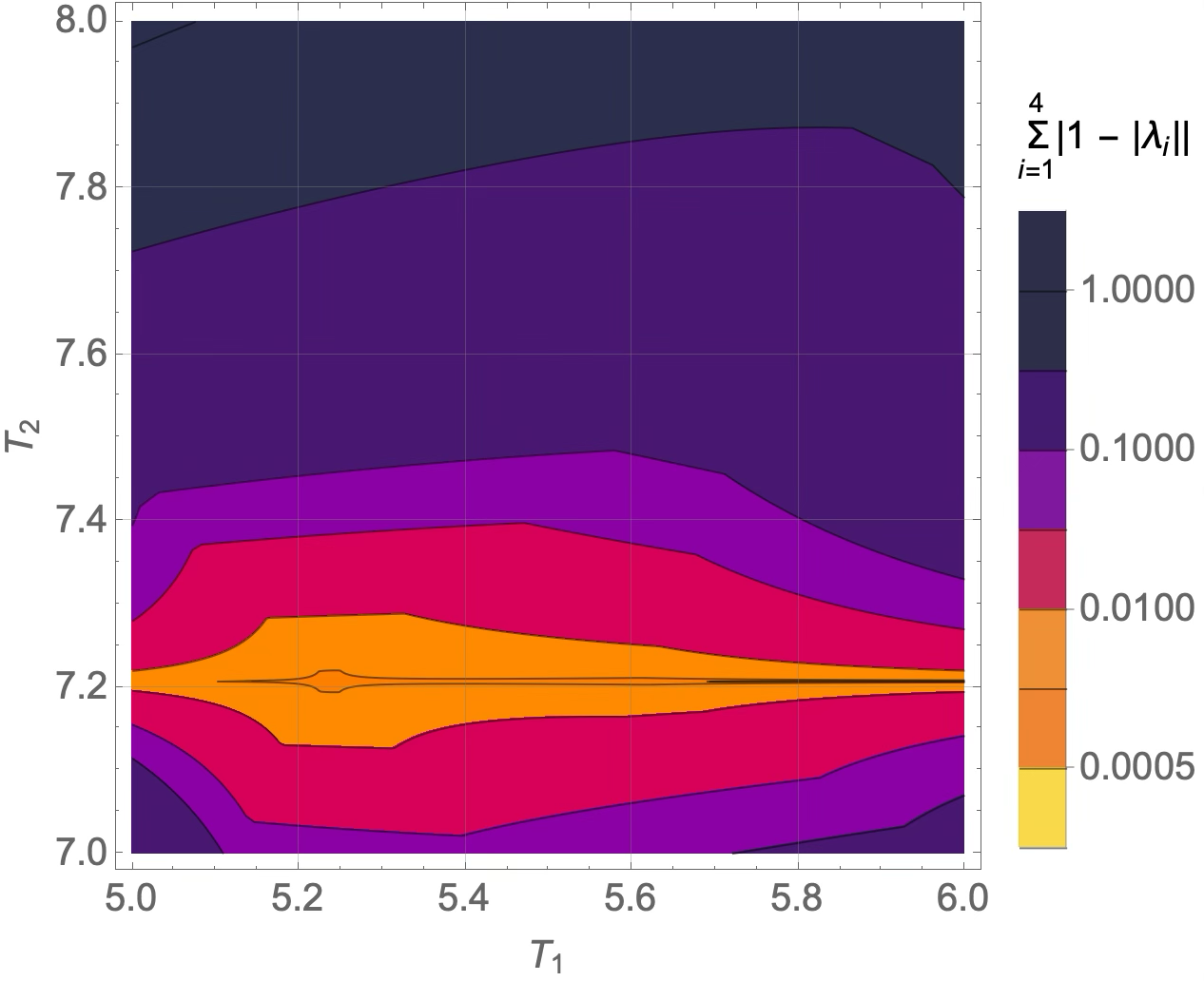}}
\caption{We have chosen $i = x, k = y, p = z$ directions {\color {black} for the tridirectional Floquet drive.} The parameters are chosen to be $\beta_1 = 0.001, \beta_2 = 0.8, \beta_3 = 0.9, T_0 = 7$. The contours are for the quantity $\sum\limits_{i=1}^4 | 1 - |\lambda_i ||$, which should vanish when all the eigenvalues are pure phases. {\bf Left :} A larger area of the $T_1-T_2$ plane with varying values of the eigenvalues. We zoom into the yellow rectangular region on the right plot.  {\bf Right :} Zoomed in region in the $T_1-T_2$ plane with higher resolution of $\sum\limits_{i=1}^4 | 1 - |\lambda_i ||$. }\label{fig4}
\end{figure}

The $4\times 4$ analog of the matrix $V_n$ of \eqref{6-4} now has entries the depend on 6 parameters. Though the number of parameters have increased, so does the number of conditions for obtaining pure phase eigenvalues. Numerically, for generic choices of parameters, we do not find any region where all the four eigenvalues ($\lambda_i$) are pure phases, as Fig.\ref{fig4} indicates. Therefore the system does not seem to reach any {\em elliptic} steady state in the 6 parameter space of $\beta_i, T_i$. We emphasize that we have not been able to prove this analytically but checked numerically in various extreme limits in the parameter space.  In the left panel of Fig.\ \ref{fig4}, where one of the deformations, $\beta_1$, is chosen to be very small compared to the other two, while an elliptic phase develops at $\beta_1 = 0$ (the two directional case) we do not see any elliptic region opening up in the parameter space.

\section{Perturbative computation of Floquet Hamiltonians}
\label{pert} 

In general, it is difficult to obtain an explicit expression for the Floquet Hamiltonian for a multi-step drive.\footnote{In fact, it is not expected to be described by a local expression for low drive frequencies.} However, they may be computed using 
perturbative schemes \cite{rev11}. In this work, we shall two such schemes. The first constitutes a perturbative expansion in the high frequency regime; this expansion is termed as the Magnus expansion \cite{rev9,rev11}. 
The second involves an expansion in the high drive amplitude regime and is termed as FPT \cite{dsen1,nigel1,rev11}. We compute the Floquet Hamiltonians using such schemes for multi-step drives and discuss the corresponding 
Killing horizons. It is well-known that the leading term of these Floquet Hamiltonians reproduces the evolution of the driven system quite 
accurately up to a large prethermal time window in the high frequency for amplitude regime \cite{saito1}; hence we expect our analysis to be valid at sufficiently long times in both cases.

\subsection{Magnus Expansion}

In the high-frequency regime, {\it i.e.}~when the drive frequency $\omega_D = (2\pi)/T \gg 1$, a perturbative expression can be written down in powers of $T$ \cite{rev9}. This series expansion is known as the Magnus expansion 
and it provides an approximate description of the dynamics up to the time-scale ${\cal O} (e^{C \omega_D})$, where $C$ is an order-one constant \cite{saito1}. Formally, the expansion takes the following form (for details see App.\ \ref{Magnusapp})    
\begin{eqnarray}
    H_F \equiv \sum_{n=0}^\infty H_F^{(n)} = \sum_{n=1}^\infty T^{n-1} \Omega_n \ ,
\end{eqnarray}
where 
\begin{eqnarray}
   && \Omega_1 = \frac{1}{T} \int_0^T dt H(t) \ , \\
   && \Omega_2 = \frac{1}{2i \hbar T^2} \int_0^T dt_1 \int_0^{t_1} dt_2 \left[H(t_1), H(t_2) \right] \ ,   
\end{eqnarray}
where $H(t)$ is the time-dependent Hamiltonian. These expressions yield:
\begin{eqnarray}
&& H_F^{(1)} = \Omega_1 = \frac{1}{T} \int_0^T dt H(t) \ , \\
&& H_F^{(2)} = T \Omega_2 = \frac{1}{2i \hbar T} \int_0^T dt_1 \int_0^{t_1} dt_2 \left[H(t_1), H(t_2) \right] \ .
\end{eqnarray}

Let us now choose the following time-dependent Hamiltonian:
\begin{eqnarray}
H & = & H_1 \quad t \le T_1 \nonumber\\
& = & H_2 \quad T_1 \le t \le T \ ,
\end{eqnarray}
where 
\begin{eqnarray}
&& H_{1,2} = \alpha D + \beta_{1,2} P_{1,2} + \gamma_{1,2} K_{1,2} \ , \\
&& \beta_{1,2}^* = \gamma_{1,2} \ . 
\end{eqnarray}
The Magnus expansion for the Floquet Hamiltonian is given by
\begin{eqnarray}
H_F = \sum_{n=1}^\infty T^{n-1} \Omega_n = \sum_{n=1}^\infty H_F^{(n)} \ .
\end{eqnarray}
Under the Magnus expansion, the first order Floquet Hamiltonian is given by
\begin{eqnarray}
H_F^{(1)} = \nu H_1 + \left( 1 - \nu \right)  H_2 = \alpha D +\beta \left(\nu P_1 + \left(1 - \nu \right) P_2 \right) + \gamma \left( \nu K_1 + \left(1 - \nu \right) K_2 \right)\ , 
\end{eqnarray}
where $\nu = T_1/T$. 

At the next order in the Magnus expansion, we obtain:
\begin{eqnarray}
    H_F^{(2)} &=& \frac{T}{2 i \hbar} \nu \left( 1 - \nu \right) \left[ H_2, H_1 \right] =  \frac{T}{2} \nu \left(1 - \nu \right) \left[\alpha \left( \beta (P_1-P_2) + \gamma(K_2 - K_1)\right) + 4\beta\gamma L_{21} \right] \nonumber\\
    \label{10-1}
\end{eqnarray}  
In the next section we will use these expressions to investigate the presence of Killing horizons in the bulk.

\subsection{Floquet perturbation theory}

In this section, we consider the high amplitude expansion using FPT. To this end, we choose the following protocol: 
\begin{eqnarray}
H & = & 2 i \left(\lambda + \delta \lambda  \right) D +  i \beta_1 (K_1 + P_1 ) \ , \quad t \le T/2 \ , \\
& = & 2 i \left(- \lambda + \delta \lambda  \right) D +  i \beta_2 (K_2 + P_2 ) \ , \quad t > T/2 \ .  \label{protfpt1} 
\end{eqnarray}
with $\lambda \gg \delta \lambda, \beta_1,\beta_2$. With this choice, the parameters $\{\lambda, \delta \lambda, \beta_{1,2}\}$ are all real-valued.\footnote{Note that, it follows: 
\begin{eqnarray}
    H^\dagger = H \quad \implies \quad D^\dagger = - D \ , \quad  K^\dagger = - P \ , \quad P^\dagger = - K \ .
\end{eqnarray}
This is the same adjoint action as noted in (\ref{adjact}).}

For our choice of parameters, it is natural to treat the terms proportional to $\delta \lambda$, $\beta_1$, and $\beta_2$
perturbatively. To this end, we first write \cite{rev11} 
\begin{eqnarray} 
U_0(t,0) &=& e^{-2 i \lambda D t/\hbar}, \quad   t\le T/2 \nonumber\\
&=&  e^{-2i \lambda D (T-t)/\hbar},  \quad  T/2 < t\le T \label{uoeq}
\end{eqnarray}
Note that at $t=T$, $U_0(T,0)=I$ and hence $H_F^{(0)}=0$. 

The first order correction to the evolution operation may be computed as follows. We first denote $H_1(t)$ to be the perturbative 
part of the Hamiltonian which can be read off from \eqref{protfpt1} as 
\begin{eqnarray} 
H_1(t) &=&  H_1^{(1)} \theta(T/2-t) + H_1^{(2)} \theta(t-T/2), \quad  H_1^{(a)} = \delta \lambda D + i \beta_a (K_a + P_a)  \label{perthamffpt}
\end{eqnarray} 
where $\theta(x)$ denotes the step function. Using this, as elucidated in Refs.\ \cite{rev11,dsen1,nigel1}, one can obtain the first order
correction, $U_1$ to the evolution operator as 
\begin{eqnarray} 
U_1(T,0) &=& \int_0^{T/2} U_0^{\dagger}(t,0) H_1^{(1)} U_0(t,0) + \int_{T/2}^T  U_0^{\dagger}(t,0) H_1^{(2)} U_0(t,0) \label{ufirst}
\end{eqnarray} 
Since $H_1^{(1)}$ and $H_1^{(2)}$ have simple representations in terms of two different sets of $SU(1,1)$ matrices (with $D$ being the common
element in both; see \eqref{4-7}), these integrals can be easily evaluated. Writing $H_F = i \hbar U_1(T,0)/\hbar$ in this perturbative regime, one obtains the Floquet Hamiltonian up to ${\cal O}(\delta \lambda)$, ${\cal O}(\beta_{1,2})$ to be (see App.\ \ref{fptapp} for details) 
\begin{eqnarray}\label{floqamp}
 &&   H_F  = \delta \lambda \left(2 i D\right) + \frac{\sin\chi}{2\chi} \left[i \beta_1\left( K_1 e^{-i \chi} + P_1 e^{i \chi } \right)+ i \beta_2\left( K_2 e^{-i \chi} + P_2 e^{i \chi } \right) \right] \ , \nonumber \\
   && \chi = \frac{\lambda T}{2\hbar} \ . 
\end{eqnarray}
We note that for $T\to 0$, $\chi \to 0$ and $H_F$ reduces to that the obtained within the first order Magnus expansion obtained in the previous section. 
Also, we note that for $\chi= p \pi$, where $p \in Z$, $[H_F,D]=0$. This constitutes an approximate emergent symmetry of the Floquet Hamiltonian at these drive frequencies. This symmetry is emergent since it is not present in $H(t)$ for any $t$; also it is approximate since it is usually not respected by higher-order terms in $H_F$ \cite{rev12}. However, this emergent symmetry plays a key role in dynamics of the system in the prethermal regime where the first order Floquet Hamiltonian controls the dynamics. In the present case, the effect of the emergent symmetry constitutes in realization of 
non-heating dynamics in the prethermal regime at and around the special frequencies.

\section{Bulk Horizons for Floquet Hamiltonians}
\label{killing1} 

For a given periodic drive, the Floquet Hamiltonian in a $d+1$ dimensional CFT is a specific linear combination of the standard conformal generators. In a holographic theory, this vector field lifts to a isometry generator of the dual $AdS_{d+2}$. In this section we will use the approximate Floquet Hamiltonians obtained in the previous section to examine if the corresponding Killing vector field have horizons. We will restrict our attention to $d=2$, but a similar analysis can be carried out for arbitrary $d$.

\subsection{Global Lorentzian AdS$_4$}

A Lorentzian AdS$_4$ geometry can be described as a hypersurface in an $R^{2,3}$ embedding space:
\begin{eqnarray}
&& \eta_{ab}X^a X^b \equiv - (X_0)^2 - (X_4)^2 +(X_1)^2 + (X_2)^2 + (X_3)^2 = - \ell^2 \ , \\
&& \eta_{ab}= {\rm diag} (-1, 1, 1,1,  -1) \ , \quad X_a = \eta_{ab } X^b \ . 
\end{eqnarray}
The global patch is obtained by the following solution to the embedding equation:
\begin{eqnarray}
&& X^0 = -\cosh\rho \cos t \ , \quad X^4 = -\cosh\rho \sin t \ , \\
&& X^1 =  \sinh \rho \cos \theta \ , \quad X^2 = \sinh \rho \sin \theta \cos\phi \ ,  \quad X^3 = \sinh \rho \sin \theta \sin \phi  \ ,
\end{eqnarray}
which yields the following metric on AdS$_4$:
\begin{eqnarray}
ds^2 = d\rho^2   - \cosh^2\rho  dt^2 + \sinh^2 \rho \left( d\theta^2 + \sin^2\theta d\phi^2 \right) \ . \label{AdS4P}
\end{eqnarray}
The corresponding symmetry group is SO$(2,3)$ with ten generators of the following form:
\begin{eqnarray}
J_{ab} = X_a \partial_b - X_b \partial_a \ .
\end{eqnarray}

We can use the pull-back relation on the co-tangent frame:
\begin{eqnarray}
\partial_{\xi^\alpha} = \frac{\partial X^a}{\partial \xi^\alpha} \partial_{X^a} \ , \label{pullback2}
\end{eqnarray}
to express the generators in the global patch. To do so, note the following combinations that yield independent $sl(2,R)$-algebra:
\begin{eqnarray}
&& D \equiv -J_{04} \ , \quad P \equiv i \left( J_{01} + i J_{14} \right)   \ , \quad K \equiv - i \left( J_{01} - i J_{14} \right) \ , \label{confgen1}\\
&& D \equiv -J_{04} \ , \quad P \equiv  i \left( J_{03} + i J_{34} \right)  \ , \quad K \equiv - i \left( J_{03} - i J_{34} \right)   \ , \label{confgen2}\\
&& D \equiv -J_{04} \ , \quad P \equiv  i \left( J_{02} + i J_{24} \right)  \ , \quad K \equiv - i \left( J_{02} - i J_{24} \right)   \ ,\label{confgen3}
\end{eqnarray}
It is straightforward to check that each of the above combinations satisfy the $sl(2,R)$-algebra:
\begin{eqnarray}
\left[D, P \right] = - i P \ , \quad \left[D, K \right] = i K \ , \quad \left[P, K \right] = 2 i D \ . 
\end{eqnarray}
Note that, the above Killing vectors satisfy the bulk Killing equation $\nabla_a K_b + \nabla_b K_a = 0$.

Before moving forward, note the following. Let us define the adjoint action for the generators $J_{ab}$ in the following manner:
\begin{eqnarray}
    \int \psi^* \left(J_{ab} \phi \right) = \int \left( J_{ab}^\dagger \psi^* \right) \phi \ ,
\end{eqnarray}
where $\psi$ and $\phi$ are complex-valued wavefunctions. This definition yields:
\begin{eqnarray}
    J_{ab}^\dagger = - J_{ab} \ . 
\end{eqnarray}
This yields the following adjoint action on the conformal generators:
\begin{eqnarray}
    D^\dagger = - D \ , \quad P^\dagger = - K \ , \quad K^\dagger = - P \ . \label{adjact}
\end{eqnarray}

In what follows, we shall use these results for obtaining Killing horizons for periodic drive protocols.

\subsection{Killing Horizons in the Magnus Expansion}
%
To lowest order in the Magnus expansion, we can now identify the corresponding bulk Killing vector as:
\begin{eqnarray}
H_F^{(0)} = \xi^A \partial_A \ , 
\end{eqnarray}
For concreteness, let us now choose:
\begin{eqnarray}
&& P_1 = i (J_{03} + i J_{34}) \ , \quad K_1 = - i (J_{03} - i J_{34}) \ , \\
&& P_2 = i (J_{02} + i J_{24}) \ , \quad K_2 = - i (J_{02} - i J_{24}) \ . 
\end{eqnarray}
It is now possible to explicitly construct the norm of the Killing vector, however, it is not particularly illuminating to do so.

Note that, the algebraic structure of the vanishing Killing norm equation is given by
\begin{eqnarray}
    A + B Y^2 - Y^2\left(C + D \frac{\sqrt{Y^2-1}}{Y} \right)^2 = 0 \ , \label{Mag1_Kill}
\end{eqnarray}
where $A,B,C,D$ are $Y$-independent and $Y = \cosh\rho$. This algebraic equation is already suggestive in the $Y\gg 1$ limit, in which one obtains a purely quadratic equation, with only a non-extremal solution:
\begin{eqnarray}
    Y^2 = \frac{A}{\left(C + D \right)^2 - B} \ .
\end{eqnarray}

The general solution of equation (\ref{Mag1_Kill}) takes the schematic form:
\begin{eqnarray}
    Y^2 = \frac{\pm 2 \sqrt{C^2 D^2 \left(A
   \left(A+B-C^2\right)+D^2
   (A+B)\right)}+\left(A+D^2\right)
   \left(D^2-B\right)+C^2
   \left(A-D^2\right)}{-2D^2 \left(B+C^2\right)+\left(B-C^2\right)^2+D^4} \ , \nonumber\\
\end{eqnarray}
which implies clearly that the necessary condition for a real solution is: 
\begin{eqnarray}
{\mathcal D} &=& C^2 D^2 \left(A \left(A+B-C^2\right)+D^2(A+B)\right) \ge 0 \ , \label{disc}
\end{eqnarray}
where the equality corresponds to an extremal solution. For the lack of a better word, let us refer to ${\mathcal D}$ as {\it discriminant}.

It is somewhat unwieldily, but possible to write down the expressions of $A,B,C,D$ explicitly, which yields:
\begin{eqnarray}
&&    A = - e^{-2 i t} \sin ^2(\theta ) \left((\nu-1\cos\phi \left(\beta_2-\gamma_2 e^{2 i
   t}\right)- \nu  \sin \phi \left(\beta_1-\gamma_1 e^{2 i t}\right)\right)^2 \ , \nonumber\\
&& B = \cos ^2\theta  (\sin t (\nu  (\beta_1 + \gamma_1) \sin \phi + (\nu -1)
   (-\beta_2 - \gamma_2) \cos \phi)-\nonumber\\
   && i
   \cos t (\nu  (\gamma_1 - \beta_1)
   \sin \phi + (\nu -1) (\beta_2 - \gamma_2) \cos \phi))^2+ \nonumber\\
   && (\sin t (\nu  (\beta_1 + \gamma_1) \cos \phi +(\nu -1) (\beta_2+\gamma_2) \sin \phi)+ \nonumber\\
   && i
   \cos t (\nu  (\beta_1 - \gamma_1)
   \cos \phi +(\nu -1) (\beta_2 - \gamma_2) \sin \phi))^2   \ , \\
   && D= e^{-i t} \sin \theta  \left(\nu  \sin \phi 
   \left(\beta_1+\gamma_1 e^{2 i
   t}\right)+(\nu -1) \cos \phi  \left(-\beta_2-\gamma_2 e^{2 i t}\right)\right) \ , \\
   && C = \alpha \ . 
\end{eqnarray}
With these, the discriminant can also be written down explicitly, however, this is a very long expression and is not very illuminating. We would expect, if the presence of a horizon is coupled to the type of conjugacy classes of the corresponding
transformations, then a purely algebraic condition can be obtained. It is possible to check that the following general case:
\begin{eqnarray}
    && \beta_1 = \beta \ , \quad \gamma_1 = \gamma \ , \\
    && \beta_2 = s \beta \ , \quad \gamma_2 = s \gamma \ , \quad s \in R \ ,
\end{eqnarray}
one obtains
\begin{eqnarray}
{\mathcal D} = & \alpha ^2 e^{-4 i t} \sin ^4\theta \left(\beta ^2-\gamma ^2 e^{4 i t}\right)^2 \left(\alpha ^2-4 \beta  \gamma  \left(\nu ^2+(\nu
   -1)^2 s^2\right)\right) \nonumber\\
 &  ((\nu -1) s \cos \phi -\nu  \sin \phi )^4   \ . \label{discMagnus1}
\end{eqnarray}
It is clear from (\ref{discMagnus1}) that, indeed, we obtain a purely algebraic condition:
\begin{eqnarray}
&&  \left(\alpha ^2-4 \beta  \gamma  \left(\nu ^2+(\nu-1)^2 s^2\right)\right) = 0 \ , \\
&& \implies \quad \alpha^2 = 4 |\beta|^2 \left(\nu ^2+(\nu-1)^2 s^2\right) \ . \label{cond_Mag1}
\end{eqnarray}
The two limiting cases are easy to check: (i) $\nu=0$ yields: $\alpha^2 = 4 s^2 |\beta|^2$, which is equivalent to the case of $H_F = H_2$. (ii) Similarly, $\nu=1$ yields: $\alpha^2 = 4 |\beta|^2$, which is equivalent to $H_F = H_1$ as expected. 

It is important to note that when $s\in {\mathbb C}$, then no such purely algebraic relation exists. The discriminant, in this case, always intertwines space-time coordinates with the parameters of the problem. It indicates that, in such a case, the existence of the horizon may not be captured with a simple algebraic relation. 

Consider now the next order in the Magnus expansion, equation (\ref{10-1}).
We will set $\hbar=1$ here on. Up to this order, we can again explicitly write down the expression for the Killing norm. This yields:
\begin{eqnarray}
    ||\xi ||^2 = A+B Y^2+Y^2\left(\left(C+\frac{D\sqrt{Y^2-1}}{Y}\right)^2\right)+2 E Y\sqrt{Y^2-1} \ , \label{Killing_norm_M2}
\end{eqnarray}
where
\begin{eqnarray}
A &=& \alpha  (\nu -1) \nu  T \sin ^2\theta \sin (2 \phi ) (\beta_2 \gamma_1 - \beta_1 \gamma_2)-e^{-2 i
   t} \sin ^2\theta  \left((\nu -1) \cos \phi  \left(\beta_2-\gamma_2 e^{2 i t}\right) \right.\nonumber\\
   & - & \left. \nu  \sin \phi 
   \left(\beta_1 - \gamma_1 e^{2 i t}\right)\right)^2 \ ,
\end{eqnarray}
\begin{eqnarray}
B & =& \cos ^2\theta  (\sin t (\nu  (\beta_1+\gamma_1)
   \sin \phi+(\nu -1) (-\beta_2-\gamma_2) \cos\phi) \nonumber\\
  &-&  i \cos t (\nu (\gamma_1-\beta_1)\sin \phi+(\nu -1) (\beta_2-\gamma_2) \cos\phi))^2 \nonumber\\
  &+&(\sin t (\nu (\beta_1+\gamma_1)\cos\phi+(\nu -1) (\beta_2+\gamma_2) \sin\phi)\nonumber\\
  & + & i \cos t (\nu  (\beta_1-\gamma_1)\cos\phi+(\nu -1) (\beta_2-\gamma_2) \sin\phi))^2 \nonumber\\
  & + &\alpha  (\nu -1) \nu  e^{-2 i t} T \left(\beta_1^2 \nu +\beta_2^2 (\nu -1)-e^{2 i t} \sin ^2\theta
   \sin (2 \phi ) (\beta_2 \gamma_1-\beta_1 \gamma_2) \right. \nonumber\\
   & + & \left. e^{4 i t} \left(\gamma_2^2-\nu \left(\gamma_1^2+\gamma_2^2\right)\right)\right)    \ ,
\end{eqnarray}
\begin{eqnarray}
D & = & e^{-i t} \sin\theta \left(\nu  \sin \phi \left(\beta_1+\gamma_1 e^{2 i t}\right)+(\nu -1) \cos \phi\left(-\beta_2-\gamma_2 e^{2 i t}\right)\right)    
\end{eqnarray}
\begin{eqnarray}
E & = & \frac{1}{2} (\nu -1) \nu  e^{-i t} T \sin \theta \left(\sin\phi \left(\alpha ^2 \left(\beta_1 - \gamma_1 e^{2 it}\right)+2 (\nu -1) \left(\beta_2-\gamma_2 e^{2i t}\right) (\beta_1 \gamma_2+\beta_2\gamma_1)\right) \right. \nonumber\\
& + & \left. \cos\phi \left(\alpha ^2\left(-\beta_2+\gamma_2 e^{2 i t}\right)+2 \nu \left(\beta_1-\gamma_1 e^{2 i t}\right)(\beta_1\gamma_2+\beta_2 \gamma_1)\right)\right) \ ,
\end{eqnarray}
\begin{eqnarray}
C & = & \alpha \ .
\end{eqnarray}

As before, in the limit $Y\gg 1$, equation (\ref{Killing_norm_M2}) becomes a purely quadratic equation, that admits only a non-extremal solution:\footnote{Physically, this simply implies that no extremal horizon exists when $Y\gg 1$.}
\begin{eqnarray}
    Y^2 = \frac{A}{\left( C + D \right)^2 -B - 2 E} \ .
\end{eqnarray}

It is also possible to explicitly solve the $|| \xi ||^2 = 0$ equation using (\ref{Killing_norm_M2}), which yields:
\begin{eqnarray}
Y^2 & = & \frac{1}{B^2-2B\left(C^2+D^2\right)+\left(C^2-D^2\right)^2+8 CD E - 4E^2} \nonumber\\
& & \left[ \pm 2 \sqrt{(E-CD)^2\left(A^2+A\left(B-C^2+D^2\right)+BD^2-2CDE+E^2\right)} \right. \nonumber\\
&+ & \left.A\left(-B+C^2+D^2\right)+D^2\left(-B-C^2\right)+4 C D E+D^4-2E^2 \right] \ .   
\end{eqnarray}
Clearly, there is again an emerging {\it discriminant} that defines an extremal horizon. It is not illuminating to explicitly write down how the purely algebraic condition emerges in this case as well, however, it is straightforward to check that, with:
\begin{eqnarray}
     && \beta_1 = \beta \ , \quad \gamma_1 = \gamma \ , \\
    && \beta_2 = s \beta \ , \quad \gamma_2 = s \gamma \ , \quad s \in R \ ,
\end{eqnarray}
one obtains:
\begin{eqnarray}
\left[\alpha^2 - 4 |\beta|^2 \left( \nu^2 + s^2 \left(1-\nu \right)^2\right)  \right] \left( {\rm others}\right) = 0 \ , \label{eq:killing}
\end{eqnarray}
which is an identical condition to (\ref{cond_Mag1}).The higher order terms in Magnus expansion may also be derived systematically. However, they lead to cumbersome expressions of the Killing norm which are not particularly illuminating; we do not explore them here.

\noindent
Note that from the quaternionic analysis of the bidirectional drive the exact phase boundary was given by \eqref{eq:2stepbndy}. The Magnus regime corresponds to the high frequency driving where both $T_1, T_2$ are small. Expanding Eq.\eqref{eq:2stepbndy} with $T_1 = \nu T, T_2 = T, \beta_1 = \beta$ and $\beta_0 = s \, \beta$ we get: 
\begin{align}
1 + \frac{1}{2} T^2 \bigg( -1 + \beta^2 \left( \nu^2 + s^2 ( \nu - 1)^2 \right)\bigg) + {\cal O}(T^4) &= 1.
\end{align}
When the L.H.S is truncated till quadratic order we get: 
$$
 \beta^2 \left( \nu^2 + s^2 ( \nu - 1)^2 \right)= 1. 
 $$
 In Eq.\eqref{eq:killing} if we {\color{ black}plug in} $\alpha =2$ which is the case in our field theory analysis, then it matches exactly with the above equation.

\subsection{Bulk Horizons in Floquet Perturbation Theory}

In this subsection we investigate bulk horizons which follow from the Floquet Hamiltonian in Floquet perturbation theory.

As before, we can express the Floquet Hamiltonian in terms of the bulk Killing vectors: $H_F^{(1)} = \xi^A \partial_A$, where the conformal generators are given in (\ref{confgen1})-(\ref{confgen3}). The algebraic structure of the vanishing Killing norm equation, as before, takes the following form:
\begin{eqnarray}
A + B Y^2 - Y^2\left(C + D \frac{\sqrt{Y^2-1}}{Y} \right)^2 = 0 \ ,
\end{eqnarray}
where
\begin{eqnarray}
 &&    A = \frac{1}{4\chi^2}4 \sin ^2\theta \sin^2\chi \sin ^2(t-\chi) (\beta_1 \sin (\phi )+\beta_2 \cos (\phi ))^2 \ , \\
 && B = \frac{1}{4\chi^2}4 \sin ^2\chi \sin ^2(t-\chi) \left(\cos ^2\theta (\beta_1 \sin \phi+\beta_2 \cos \phi)^2+(\beta_1\cos \phi-\beta_2
   \sin \phi)^2\right)  \nonumber \\
 &&   C = \frac{1}{2\chi}4 \delta \lambda  \chi \ , \\
 && D = \frac{1}{2\chi}2 \sin\theta \sin \chi \cos (t-\chi) (\beta_1 \sin \phi +\beta_2\cos\phi) \ . 
\end{eqnarray}

\begin{figure}
\centering
\rotatebox{0}{\includegraphics*[width= 0.65 \linewidth]{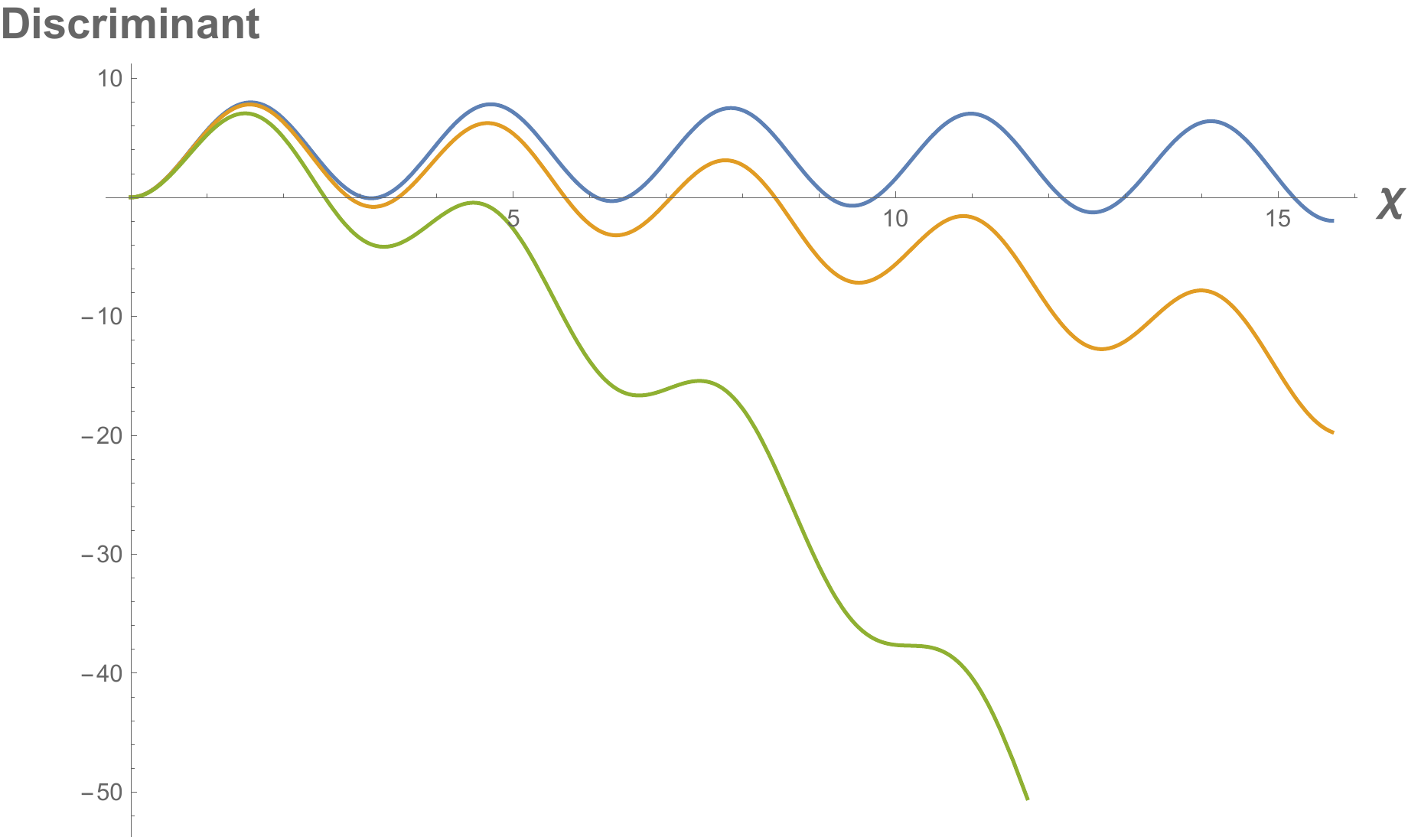}}
\caption{ We have shown here the functional behaviour of the condition in (\ref{extremeamp}). Non-extremal horizons exist when the curve takes positive values, for negative values there are no real solution for a horizon, whereas extremal horizons exist when the curve crosses zero. We have fixed $\beta^2=2$. From top to bottom, we have assigned: $\delta \lambda = \sqrt{0.001}, \sqrt{0.01}, \sqrt{0.05}$, respectively. Evidently, increasing $\delta\lambda$ reduces the parametric region that yields a horizon. We can also explicitly observe the re-entrant phase structure.  \label{fig-ampexp} }
\end{figure}

\begin{figure}
\centering
\rotatebox{0}{\includegraphics*[width= 0.65 \linewidth]{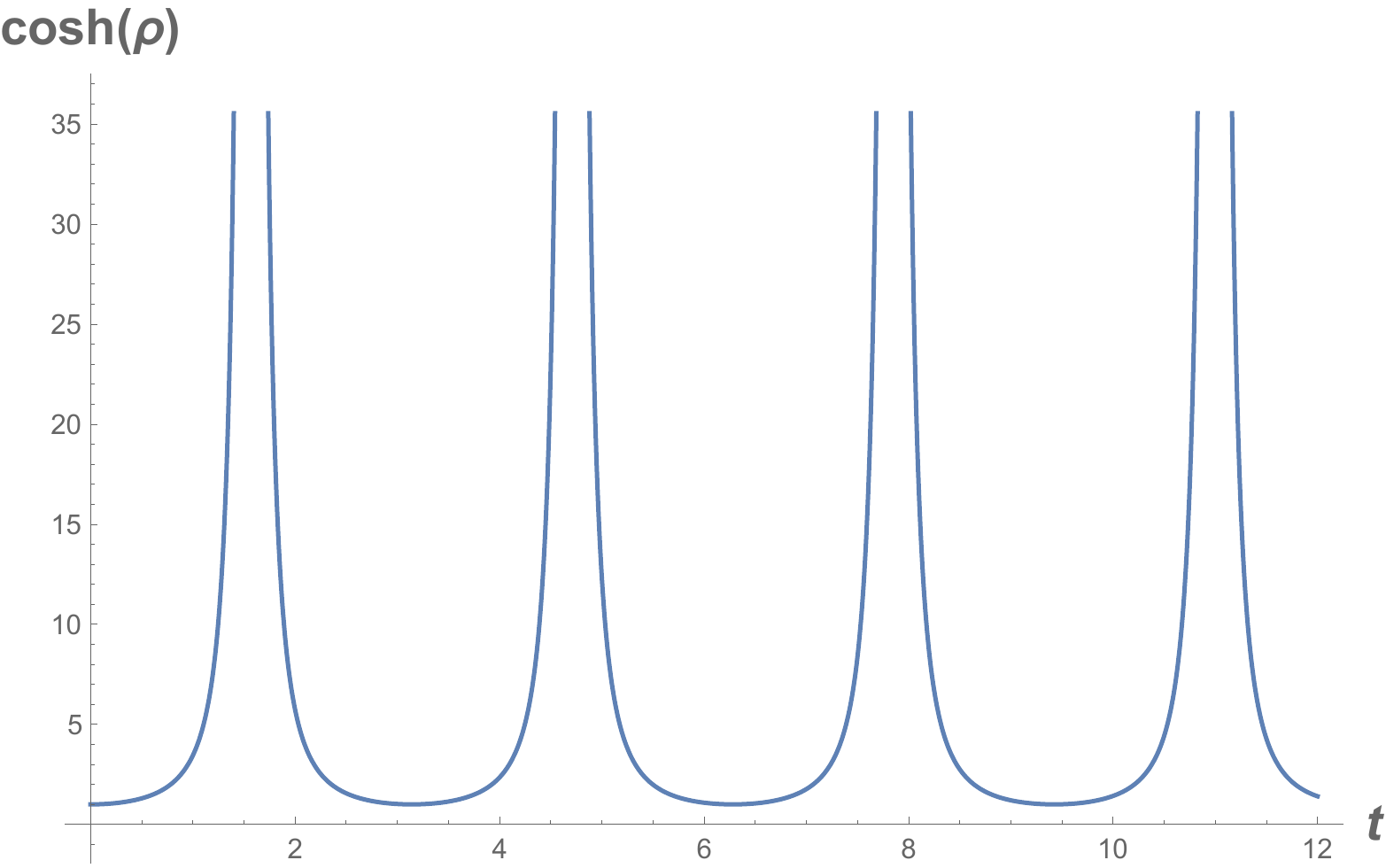}}
\caption{ We have shown here the behaviour of the horizon, in particular, $\cosh\rho$ as a function of time, $t$. We have chosen: $\beta=1$, $\delta\lambda=\frac{\sqrt{2}}{\pi}$, and for a fixed $x=\frac{\pi}{2}$ slice. The bulk horizon grows unbounded and reaches the conformal boundary periodically in time.  \label{fig-ampexphor}}
\end{figure}

As before, the necessary condition for finding real root is given by
\begin{eqnarray}
    C^2 D^2 \left(A \left( A + B - C^2\right) + \left(A+B \right) D^2\right) \ge 0 \ ,
\end{eqnarray}
while the equality yields an extremal solution. This yields:
\begin{eqnarray}\label{extremecondchi}
   &&-\frac{\delta \lambda ^2 \sin^4\theta \sin ^4\chi \sin ^2(2
   (t-\chi)) (\beta_1 \sin\phi +\beta_2 \cos\phi)^4 \left(-\beta_1^2-\beta_2^2+8 \delta
   \lambda ^2 \chi^2+\left(\beta_1^2+\beta_2^2\right)
   \cos (2\chi)\right)}{2\chi^6} \ge 0 \ . \nonumber\\
\end{eqnarray}
At the leading order in the $\chi\to 0$ limit, we obtain the condition:
\begin{eqnarray}
    \delta\lambda^2 \left(\beta_1^2 + \beta_2^2 - 4 \delta \lambda^2 \right) \sin ^2(2t) \sin^4\theta (\beta_1 \sin\phi +\beta_2 \cos\phi)^4 \ge 0 \ , 
\end{eqnarray}
which matches with the first order Magnus expansion result. In this case, the extremality condition, $\beta_1^2 + \beta_2^2 - 4 \delta \lambda^2=0$ is independent of the time-period $T$. On the other hand, the extremality condition from (\ref{extremecondchi}) is given by
\begin{eqnarray}
    \left(-\beta_1^2-\beta_2^2+8 \delta
   \lambda ^2 \chi^2+\left(\beta_1^2+\beta_2^2\right)
   \cos (2\chi)\right) = 0 \ , \label{extremeamp}
\end{eqnarray}
which depends explicitly on the time-period $T$. A pictorial representation of how this condition behaves as a function of the time-period is given in Fig.\  \ref{fig-ampexp}. From this, we observe a re-entrant behaviour; as $\chi$ is varied, the heating and non-heating phases oscillate as a function of $\chi$. We have also demonstrated how a non-extremal horizon behaves as a function of time, for a constant $x$-slice, in Fig.\ \ref{fig-ampexphor}. 

Note that, the Floquet Hamiltonian in (\ref{floqamp}) reduces to simple dilatation ($D$) when $\chi = m \pi$, $m \in {\mathbb Z}$. In these cases, $\cos(2\chi) =1$ and therefore the LHS of the extremality condition becomes: $8 \chi^2 \delta\lambda^2 $, which is non-vanishing for any non-vanishing $\chi$. It is easy to check that, the condition in (\ref{extremecondchi}) can never be satisfied and therefore we cannot have a real-valued Killing horizon here. Thus, at $\chi = m\pi$ (and $m \not=0$), we recover the physics of the non-heating phase.

Alternatively, consider the condition in (\ref{extremeamp}). Suppose $\beta_1^2+\beta_2^2=\beta^2$, then the extremality condition is satisfied at:
\begin{eqnarray}
\delta\lambda^2 = \frac{\beta^2}{4} \frac{ \sin^2\chi}{\chi^2} \ . \label{chisol}
\end{eqnarray}
Therefore, given a $\delta \lambda$, $\chi$ can be varied (or {\it vice versa}) to move through the phases. This movement is completely frozen at $\chi = m \pi$ (and $m \not = 0$), where the Killing horizon disappears. Also, since $|\sin^2\chi/\chi^2| \le 1$, the extremality condition can only be satisfied for $4\delta\lambda^2\le \beta^2$. Said differently, there is only non-heating phase, when $4 \delta\lambda^2 - \beta^2 >0 $. Note that, if we consider the following auxiliary Floquet Hamiltonian:
\begin{eqnarray}
    H_F^{\rm aux} = \left(2 \delta \lambda\right) i D + \frac{i \beta}{2} \left(P + K \right) \ ,
\end{eqnarray}
the corresponding non-heating phases will coincide with the same algebraic regime: $4 \delta\lambda^2 - \beta^2 >0$.

Let us view the solutions in (\ref{chisol}) from another perspective. Evidently, we can rewrite (\ref{chisol}) as:
\begin{eqnarray}
 \frac{ \sin^2\chi}{\chi^2} = \frac{4 \delta\lambda^2}{\beta^2} \ . \label{chisol2}
\end{eqnarray}
It is important to note that there are multiple roots of this equation. Using (\ref{chisol2}), we obtain:
\begin{eqnarray}
    && \frac{\partial \chi}{\partial \delta \lambda} = \frac{2 \chi}{2 \delta\lambda - \sqrt{\beta^2 - 4 \chi^2 \delta\lambda^2}} \ , \\
    && \frac{\partial \chi}{\partial \beta} = \frac{2 \chi \delta\lambda}{2 \beta \delta\lambda - \beta \sqrt{\beta^2 - 4 \chi^2 \delta\lambda^2}} \ . 
\end{eqnarray}
In obtaining the above, we have explicitly used the relation in (\ref{chisol2}). It is clear that, given a $\chi$, whenever the denominator vanishes, the derivatives above diverge. We can interpret this as the cases where the roots of the equation (\ref{chisol2}) become unstable with respect to a small variation in the parameters. The vanishing denominator condition yields:
\begin{eqnarray}
    2 \delta \lambda = \sqrt{\beta^2 - 4 \chi_{\rm crit}^2 \delta\lambda^2} \ . \label{rootunstable}
\end{eqnarray}
Using (\ref{chisol2}) and (\ref{rootunstable}), we obtain the critical values of $\chi$, denoted above by $\chi_{\rm crit}$, for which the roots become unstable:
\begin{eqnarray}
    &&\sin^2\chi_{\rm crit} = 1 - \frac{4 \delta\lambda^2}{\beta^2} \ . \\
    && \implies \quad \sin^2 \left(\sqrt{\frac{\beta^2}{4\delta\lambda^2} -1} \right) = 1 - \frac{4 \delta\lambda^2}{\beta^2} \ .
\end{eqnarray}
Note that, within the range $\beta^2 \ge 4 \delta \lambda^2$, there exists such critical unstable points. For example, when $4 \delta \lambda^2 \to \beta^2$, we obtain:
\begin{eqnarray}
    \chi_{\rm crit} \to  m \pi \ , \quad m \in {\mathbb Z} \quad {\rm and} \quad m \not = 0\ .
\end{eqnarray}
However, all the other roots of (\ref{chisol2}) remain stable under the variation of the parameters. Thus we find that in the large drive amplitude regime, the formation and 
disappearance of the Killing horizons can be controlled by tuning the drive frequencies.

\section{Discussion}
\label{diss} 
This paper extends our previous work on Floquet dynamics in higher-dimensional CFTs \cite{ddks1} by exploring more general drive protocols and their resulting dynamical phases.
We determined these phases by directly calculating the eigenvalues of the matrices which represent the time evolution at the end of a single cycle in terms of quaternions. These eigenvalues should also determine the type of conjugacy classes of the corresponding group element. 

Conformal transformations in four Euclidean dimensions form the group $SL(2,H)$: the type of conjugacy classes of this group are known in terms of their quaternionic representations \cite{parker-short}. 
A unique feature in higher dimensions in comparison with two, is the possibility of considering a drive in multiple directions sequentially. This can result in a situation where some entries of the evolution matrix grow exponentially with the stroboscopic time, while some others oscillate (e.g. item \ref{item4} in \S\ref{3-step-1}). {\color{ black} This leads to a hybrid phase where the stroboscopic response as measured by the unequal time correlation function has oscillations on top of an exponential decay. Our results also indicate that for a generic drive a purely oscillatory phase does not exist.} A complete characterization of these dynamical phases will require developing a proper classification framework based on quaternionic  invariants of the Lorentzian conformal group $SO(d,2)$, rather than relying on the Euclidean counterparts. {\color {black} In this work, we have taken the first step towards such a classification.} 

{\color{ black} A key result of this work is 
of this work is the establishment of a direct geometric correspondence between the dynamical phases of driven CFTs and the presence or absence of Killing horizons of the Hamiltonian vector field. We showed that heating phases correspond precisely to the presence of non-extremal Killing horizons with finite surface gravity on the base space-time of the CFT. This geometric picture provides profound insight into the mechanism of heating in driven conformal systems, revealing it as a generalized Unruh effect. For the non-heating phase there is no such horizon, while for the critical case the horizon is extremal. For holographic CFT's these horizons have extensions into the bulk $AdS$ space-time.  This is not merely an analogy but a fundamental connection arising from the shared symmetry structure between CFTs and AdS geometry. For a CFT with a single non-standard Hamiltonian, these are horizons of the bulk vector field which represents this time evolution in the bulk.}

{\color{ black}It is worth restating, that the AdS geometric description provides us with a kinematic framework for {\it all} CFTs, and for holographic CFTs, it provides us with a foundational framework for further research. For example, it is natural to think of relevant deformations of ${\mathcal N}=4$ SYM and the subsequent interplay between the deformation parameter with the choices of the CFT Hamiltonian. For such cases, the geometric description may be the only pragmatic way to address questions at strong coupling.} 

Through quaternionic representations of conformal transformations, we've analyzed the stroboscopic response of systems subjected to various multi-step drive protocols, demonstrating distinct phases ranging from purely oscillatory behaviour to exponential decay and critical power-law responses. The perturbative methods we've developed—Magnus expansion for high-frequency drives and Floquet perturbation theory for high-amplitude drives—offer analytical tools to study these systems and reveal how drive parameters can be tuned to control the formation and properties of horizons.

While our geometric considerations are based on analyzing bulk Killing horizons that become CFT generators on the boundary, we do not directly address the question of bulk dynamics of the putative dual. This goes back to the fact that in the stroboscopic dynamics, the time enters as a parameter in the conformal map\footnote{These are the $T_i$'s mentioned just ahead of Eq.\eqref{1-2}.}. This map tells us where we end up at the end of a stroboscopic period without letting us see the micromotion, {\em i.e.} the dynamics in between $T_{i-1}$ to $T_i$. However this information is crucial in order to understand the exact details of the horizon dynamics, and it is our understanding that the $AdS_3/CFT_2$ analysis in \cite{nozaki1, nozaki2, deBoer, mezei, arnab2} is a step in this direction. Future work could explore how much of this generalizes to higher dimensions and if our bulk geometries are recovered when stroboscopically probed.

In this work we worked in the conformally invariant ground state, which does not change under time evolution by the Hamiltonians we considered. The effect of the drive is, however, non-trivial in unequal time correlation functions since they involve intermediate excited states, which are not invariant under this time evolution. The dynamical phases are, however, properties of the group element which describes the time evolution at the end of a single cycle. This implies that expectation values of operators in excited primary states will also exhibit these phases. In particular, it would be interesting to the study the dynamics starting from a thermal state, {\color{ black}as has been done in $1+1$ dimensions \cite{nozaki1, arnab2}.} We plan to pursue this in the future.

Even though we have extended the dynamics beyond what was considered in \cite{ddks1} allowing us to see more of the conformal group, further generalizations are possible in two directions. Firstly, instead of $D, K_\mu, P_\mu$ one can consider $K_\mu - i\, K_\nu, P_\mu + i \, P_\nu , D - i\, M_{\mu\nu}$ which also forms another $SU(1,1)$, and thus can once again be studied using similar BCH decompositions. However now some of the conformal transformations will involve Lorentz rotations, and examining how the dynamical phases get realized in this context presents an intriguing direction. Secondly, in this paper we considered periodic drives of two and three steps. There is a growing literature on aperiodic and even chaotic drives \cite{wen2, yan11, chitra2, Zhao_2022, chitra2, fang2025}. Exploring how the notion of dynamic prethermalization manifests for high dimensional critical systems remains an open question for future investigation.

\section*{Acknowledgements}
We would like to thank J. de Boer, P. Caputa and M. Nozaki for discussions. D.D. and S.R.D would like to thank the International Center for Theoretical Sciences, Tata Institute of Fundamental Research for hospitality during the workshop "Quantum Information, Quantum Field Theory and Gravity", August 2024 where some of this work was done. 
The work of S.R.D. is supported by National Science Foundation grants NSF-PHY/211673 and NSF-PHY/2410647, 
and a Jack and Linda Gill Chair Professorship. KS thanks DST for support through SERB project JCB/2021/000030. A.K. is partially supported by CEFIPRA $6304-3$, CRG$/2021/004539$ of Govt.~of India. A.K. further
acknowledges the support of Humboldt Research Fellowship for Experienced Researchers
by the Alexander von Humboldt Foundation and for the hospitality of Theoretical
Physics III, Department of Physics and Astronomy, University of Wurzburg during the
course of this work. For the current version, we are also thankful to the \texttt{Scipost} referees for pushing us to improve the clarity of our presentation.

\appendix
\section{ The Conformal Algebra }
\label{algebra}

We will consider conformal field theories on (time) $ \times S^d$ which will be obtained by Weyl transformation and analytic continuation from $R^{d+1}$.
The conformal algebra in $R^{d+1}$ is given by
\ben
[D, P_\mu] =  i P_\mu,~~~~[D, K_\mu] = -iK_\mu~~~~[K_\mu , P_\nu] = 2iD\delta_{\mu\nu} - 2 M_{\mu\nu}
\een
\ben
[ M_{\mu\nu}, K_\rho] = \delta_{\nu \rho} K_\mu - \delta_{\mu\rho} K_\nu~~~~~[M_{\mu\nu}, P_\rho]  =  \delta_{\nu \rho} P_\mu - \delta_{\mu\rho} P_\nu 
\een
\ben
[ M_{\mu\nu}, M_{\rho\sigma}] =  i \delta_{\mu \sigma} M_{\nu\rho}-i \delta_{\nu\sigma} M_{\mu\rho} + i\delta_{\nu\rho}M_{\mu\sigma} - i\delta_{\mu\rho}M_{\nu\sigma}
\label{2-1}
\een
with all other commutators vanishing. 
The hermiticity conditions are
\ben
K_\mu^\dagger = - P_\mu~~~~~~~D^\dagger = -D~~~~~~M_{\mu\nu}^\dagger = M_{\mu\nu}
\label{2-2}
\een
The action of these generators on the coordinates $x^\mu$ on $R^{d+1}$ are given by the differential operators
\bea
D & = & -i x_\mu \partial_\mu,~~~~~~P_\mu = -i\partial_\mu  \nonumber \\
K_\mu & = &-2i x_\mu (x \cdot \partial) +i r^2 \partial_\mu \nonumber \\
M_{\mu\nu} & = &-i(x_\mu \partial_\nu - x_\nu \partial_\mu)
\label{2-3}
\eea
where $r^2 = x_\mu x^\mu$. 

A Weyl transformation to $R \times S^d$ is performed by
\bea
x^\mu = e^\tau Y^\mu
\label{2-4}
\eea
where $Y^\mu$ are unit vectors on $R^{d+1}$, $Y_\mu Y^\mu = 1$ describing the $S^d$, so that the metric becomes
\ben
ds^2 = e^{2\tau} [ d\tau^2 + d \Omega_d^2 ]
\label{2-5}
\een
where $d \Omega_d^2$ is the metric on a unit $S^{d}$. One can now parametrize $Y^\mu$ by angles on $S^d$ in the usual fashion. For example, for $d=2$ we have 
\ben
Y^0 = \cos \theta~~~~~~Y^1 = \sin \theta\cos \phi~~~~~Y^2 = \sin \theta \sin \phi
\label{2-6}
\een
The differential operators which represent the action of the conformal generators on points on (time)  $ \times S^d$ can be obtained by starting with (\ref{2-3}) and performing the coordinate transformations (\ref{2-4}). 
Under this Weyl transformation a primary operator in the CFT with conformal dimension $\Delta$ transforms as
\ben
\cO(x^\mu) \rightarrow \cO(\tau,\theta_i) = e^{\tau \Delta} \cO(x^\mu)
\label{2-8}
\een
The analytic continuation $\tau = it$ leads to the Lorentzian space-time $\rm{time} \times S^d$.

{\color{ black}
\section{Explicit form of the 3 step evolution matrix}
\label{app:Vn}
In this appendix we provide the explicit expressions for the non-zero entries in the evolution matrix $V_n$ that results from the 3 step drive in two directions discussed in \S\ref{3-step-1}. 
If we define: 
\begin{equation}
	\zeta = 2\sqrt{[{\rm Re} (p_1+iq_1)]^2 - 1}\,\,\text{
and, }\,\,	\eta = 2 \sqrt{[{\rm Re} (p_1-iq_1)]^2 - 1}
\end{equation}
so that the 4 eigenvalues \eqref{8-7} are: 
\bea
\lambda_\pm &= {\rm Re} (p_1+iq_1) \pm \zeta/2, \,\, \mu_\pm =  {\rm Re} (p_1-iq_1) \pm \eta/2,
\eea
then we can express the non-zero entries as 
\begin{align}
	V_n(1,1) &= \frac{1}{2\zeta} \bigg( \lambda_-^n ( \zeta + 2 \text{Im} ( p_1 - i \, q_1 ) ) + \lambda_+^n ( \zeta - 2 \text{Im} ( p_1  - i\, q_1 ))\bigg), \\
	V_n(1,3) &= \frac{p_2 - i\, q_2}{\zeta} \left( \lambda_+^n - \lambda_-^n \right), \\
	V_n(2,2) &= \frac{1}{2\eta} \bigg( \mu_+^n ( \eta + 2i \text{Im} ( p_1 - i \, q_1 ) ) + \mu_-^n ( \eta - 2i \text{Im} ( p_1  - i\, q_1 ))\bigg), \\
	V_n(2,4) &= \frac{p_2 + i\, q_2}{\eta} \left( \mu_+^n - \mu_-^n \right), \\ 
	V_n(3,1) &= \frac{p_2^* + i\, q_2^*}{\zeta} \left( \lambda_+^n - \lambda_-^n \right), \\
	V_n(3,3) &= \frac{1}{2\zeta} \bigg( \lambda_+^n ( \zeta + 2 \text{Im} ( p_1 - i \, q_1 ) ) + \lambda_-^n ( \zeta - 2 \text{Im} ( p_1  - i\, q_1 ))\bigg), \\
	V_n(4,2) &= \frac{p_2^* - i\, q_2^*}{\eta} \left( \mu_+^n - \mu_-^n \right), \\ 
	V_n(4,4) &=  \frac{1}{2\eta} \bigg( \mu_-^n ( \eta + 2i \text{Im} ( p_1 - i \, q_1 ) ) + \mu_+^n ( \eta - 2i \text{Im} ( p_1  - i\, q_1 ))\bigg).
\end{align}
The above dependences on $n$ shows that in the situation when one of $\lambda$ or $\mu$ is complex, four entries are exponentially growing at large $n$, while the rest non-zero ones are oscillating with $n$. 
}

\section{Magnus Expansion: Derivation}
\label{Magnusapp}

Given a general evolution operator $U(t,t_0)$, which satisfies the following equation:
\begin{eqnarray}
    \frac{\partial}{\partial t} U(t,t_0) = - \frac{i}{\hbar} H(t) U(t,t_0) \ .
\end{eqnarray}
The Magnus expansion assumes an exponential representation:
\begin{eqnarray}
    U(t,t_0) = {\rm exp} \left(\Omega(t, t_0) \right)  \ , 
\end{eqnarray}
The Magnus ansatz can be recursively solved to obtain:
\begin{eqnarray}
    \Omega(t, t_0) & = & \int_{t_0}^{t} dt_1 \left( - \frac{i}{\hbar} H(t_1)\right)  \nonumber\\
    & = & + \frac{1}{2} \int_{t_0}^t dt_1 \int_{t_0}^{t_1} dt_2 \left[ \left( - \frac{i}{\hbar} H(t_1)\right), \left( - \frac{i}{\hbar} H(t_2) \right)\right] \nonumber\\
    & = & + \frac{1}{6} \int_{t_0}^t dt_1 \int_{t_0}^{t_1} dt_2 \int_{t_0}^{t_2} dt_3 \left(\left[\left( - \frac{i}{\hbar} H(t_1) \right) , \left[ \left( - \frac{i}{\hbar} H(t_2) \right), \left( - \frac{i}{\hbar} H(t_3) \right)\right]\right]  \right. \nonumber\\
  && + \left.  \left[\left( - \frac{i}{\hbar} H(t_3) \right) , \left[ \left( - \frac{i}{\hbar} H(t_2) \right), \left( - \frac{i}{\hbar} H(t_1) \right)\right]\right]  \right) + \ldots \ . 
\end{eqnarray}

Let us consider the following Hamiltonian:
\begin{eqnarray}
    H & = & H_1 \quad {\rm for} \quad t \le T_1 \ , \\
    & = & H_2 \quad {\rm for} \quad T_1 \le t \le T \ .
\end{eqnarray}
Here $T= (2\pi) / \omega_D$ is the Time-period where $\omega_D$ is the drive frequency. Consider the following choice for $H_{1,2}$:
\begin{eqnarray}
    H_{1,2} = \alpha D + \beta P_{1,2} + \gamma K_{1,2} \ .
\end{eqnarray}
Here we choose: $\alpha \in {\mathbb R}$ and $\beta, \gamma \in {\mathbb C}$ with $\beta^*= \gamma$. This choice corresponds to choosing $D^\dagger = D$, $K^\dagger = P$.

The formal expression of the Magnus expansion is written as:
\begin{eqnarray}
    H_F = \sum_{n=1}^\infty H_F^{(n)} \ ,
\end{eqnarray}
where 
\begin{eqnarray}
    H_F^{(1)} & = & \frac{1}{T} \int_0^T dt H(t) = \nu H_1 + \left(1 - \nu  \right) H_2 \nonumber\\
    & = & \alpha D +\beta \left(\nu P_1 + \left(1 - \nu \right) P_2 \right) + \gamma \left( \nu K_1 + \left(1 - \nu \right) K_2 \right) \ ,
\end{eqnarray}
and
\begin{eqnarray}
    H_F^{(2)} & = & \frac{1}{2 i T} \int_0^T dt_1 \int_0^{t_1} dt_2 \left[H(t_1), H(t_2)\right] \nonumber\\
    & = & \frac{1}{2 i T} \int_{T_1}^T dt_1 \int_0^{T_1} dt_2 \left[H_2, H_1\right] = \frac{T}{2i} \nu \left(1 - \nu  \right)  \left[H_2, H_1\right] \ . 
\end{eqnarray}
Now, using the definitions of $H_{1,2}$ and the following commutation relations:
\begin{eqnarray}
    \left[D, P_\alpha \right] = i P_\alpha \ , \quad \left[D, K_\alpha \right] = -i K_\alpha \ , \quad \left[ P_\alpha , K_\beta \right] = 2i L_{\alpha\beta} \ , 
\end{eqnarray}
we obtain:
\begin{eqnarray}
    H_F^{(2)} = \frac{T}{2} \nu \left(1 - \nu \right) \left[\alpha \left( \beta (P_1-P_2) + \gamma(K_2 - K_1)\right) + 4\beta\gamma L_{21} \right]  \ . 
\end{eqnarray}
%

\section{Floquet perturbation theory: Derivation}
\label{fptapp}
Let us now consider the following protocol:
\begin{eqnarray}
    H & = & 2 i \left( \lambda + \delta \lambda \right) D + i \beta_0 \left(K_0 + P_0 \right) \quad {\rm for} \quad t \le \frac{T}{2} \\
    & = & 2 i \left( - \lambda + \delta \lambda \right) D + i \beta_1 \left(K_1 + P_1 \right) \quad {\rm for} \quad t > \frac{T}{2} \ .
\end{eqnarray}
Let us use an explicit Pauli matrix representation of the generators above:
\begin{eqnarray}
    D \equiv \frac{i \sigma_3}{2} \ , \quad K_0 \equiv \sigma_- \ , \quad P_0 \equiv \sigma_+ \ , \quad t \le \frac{T}{2} \\
     D \equiv \frac{i \tau_3}{2} \ , \quad K_1 \equiv \tau_- \ , \quad P_1 \equiv \tau_+ \ , \quad t > \frac{T}{2} \ . 
\end{eqnarray}

At the leading order, we neglect all terms at ${\cal O}(\delta \lambda^2)$, ${\cal O}(\beta^2)$ and ${\cal O}(\beta \delta \lambda)$ terms compared to ${\cal O}( \lambda^2)$ terms. The evolution operator, therefore, is given by
\begin{eqnarray}
    U_0 & = & {\rm exp} \left( - \frac{i}{\hbar} \lambda \sigma_3 t\right) \ , \quad t \le \frac{T}{2} \ , \\
    & = & {\rm exp} \left( - \frac{i}{\hbar} \lambda \tau_3 (T-t)\right) \ , \quad t > \frac{T}{2} \ . 
\end{eqnarray}
At the next order, the evolution operator receives corrections at ${\cal O}(\delta \lambda^2)$ and ${\cal O}(\beta^2)$. This is given by, for $t \le T/2$, 
\begin{eqnarray}
    U_1 & = &  \left(- \frac{i}{\hbar} \right) \int_0^{T/2} dt U_0^\dagger \left( \delta \lambda 2 i D + i \beta_0 \left(K_0 + P_0 \right) \right)  U_0 \nonumber\\
    & = & \Pi_1 + \Pi_2 + \Pi_3 \ .
\end{eqnarray}
Since $U_0$ commutes with $D$, we readily obtain:
\begin{eqnarray}
    \Pi_1 = - \frac{i T}{2\hbar} \delta\lambda (2 i D) \ . 
\end{eqnarray}
Now we have:
\begin{eqnarray}
    \Pi_2 = \left(- \frac{i}{\hbar} \right) \int_0^{T/2} dt (i \beta_0) e^{i \lambda \sigma_3 t/\hbar} (i \sigma_-) e^{-i\lambda \sigma_3 t/\hbar} \ . 
\end{eqnarray}

To evaluate $\Pi_2$, let us introduce eigenstates of the $\sigma_3$ operator, denoted by $|m\rangle$. The matrix elements of $\Pi_2$ is given by
\begin{eqnarray}
    \langle m |\Pi_2 |n \rangle & = & \left(- \frac{i}{\hbar} \right) \delta_{m , n-1} i \int_0^{T/2} e^{2i \lambda t /\hbar} dt \nonumber\\
    & = & \left(- \frac{i}{\hbar} \right) \left(i \beta_0 \right) \delta_{m, n-1} \frac{e^{i\lambda T/\hbar}-1}{2i\lambda/\hbar} \ . 
\end{eqnarray}
Noting that $\delta_{m,n-1} = \sigma_-$, one obtains:
\begin{eqnarray}
    \Pi_2 = \left(- \frac{i}{\hbar} \right) \frac{T}{2} \left(i \beta_0 \right) \sigma_- \frac{\sin\chi}{\chi} e^{i \chi} \ , \quad \chi = \frac{\lambda T}{2\hbar} \ . 
\end{eqnarray}
Proceeding similarly, one also obtains:
\begin{eqnarray}
    \Pi_3 = \left(- \frac{i}{\hbar} \right) \frac{T}{2} \left(i \beta_0 \right) \sigma_+ \frac{\sin\chi}{\chi} e^{- i \chi} \ ,
\end{eqnarray}
Combining everything, for $t\le T/2$, we obtain:
\begin{eqnarray}
    U_1 = \left(- \frac{i}{\hbar} \right) \frac{T}{2} \left[\delta \lambda 2 i D + i \beta_0 \left(K_0 e^{-i\chi} + P_0 e^{i \chi} \right) \frac{\sin\chi}{\chi} \right]  \ ,
\end{eqnarray}
A similar calculation also yields the first order correction to the evolution operator for $t> T/2$:
\begin{eqnarray}
    U_1' = \left(- \frac{i}{\hbar} \right) \frac{T}{2} \left[\delta \lambda 2 i D + i \beta_1 \left(K_1 e^{-i\chi} + P_1 e^{i \chi} \right) \frac{\sin\chi}{\chi} \right]  \ ,
\end{eqnarray}

Therefore, at the first sub-leading order, the complete evolution operator is given by
\begin{eqnarray}
e^{-i H_F^{(1)} T/\hbar} \equiv U_1 = U_1 + U_1' \ ,
\end{eqnarray}
which yields:
\begin{eqnarray}\label{ampHF}
&& H_F^{(1)} = \delta\lambda \left(2 i D \right) + \frac{\sin\chi}{\chi} \left[\frac{\beta_0}{2} \left( K_0 e^{-i \chi} + P_0 e^{i\chi}\right) + \frac{\beta_1}{2} \left( K_1 e^{-i\chi} + P_1 e^{i \chi}\right)  \right]  \ , \nonumber\\
&& \chi = \frac{\lambda T}{2\hbar} \ .
\end{eqnarray}
The above result is valid at the first sub-leading order in an expansion where the perturbation parameter is ${\cal O}\left( \delta \lambda/\lambda\right)$ and ${\cal O}\left(\beta_{0,1}/\lambda \right)$. The higher order corrections are therefore suppressed in powers of $\lambda^{-1}$. Note that,

It is instructive to consider the $\chi \to 0$ limit, where $\sin\chi/\chi \to 1$ and the effective Floquet Hamiltonian in (\ref{ampHF}) reduces to the first order Magnus expansion result. On the other hand, at $\chi = m\pi$, where $m\in {\mathbb Z}$ and $m\not = 0$, $H_F^{(1)} \sim D$, which corresponds to the deep end of the non-heating phase. As one varies $\chi$, it is now possible to navigate through the various phases.

\section{Killing vectors in Poincar\'{e} AdS}
\label{Poincare}

Let us analyze the structure of the Killing horizon in the Poincar\'{e} patch in AdS. First, we will discuss the AdS$_3$ geometry, which has the following metric:
\begin{eqnarray}
ds^2 = \ell^2 \left( \frac{du^2}{u^2} + u^2 \left( - dt^2 + dx^2 \right) \right) \ . 
\end{eqnarray}
Correspondingly, the six explicit Killing vectors are given by
\begin{eqnarray}
&& J_{01} = \frac{t x}{\ell} \partial_t  - \frac{u x}{\ell} \partial_u + \frac{u^2 \left(- \ell^2 + t^2 + x^2\right)-1}{2 \ell^2 u^2} \partial_x \ , \\
&& J_{02} = - t \partial_t  + u \partial_ u - x \partial_x \ , \\
&& J_{03} = -\frac{u^2 \left( \ell^2 + t^2 + x^2\right) +1}{2 \ell^2 u^2} \partial_t  + \frac{t u}{\ell} \partial_u - \frac{t x}{\ell} \partial_x \ , \\
&& J_{12} = \frac{t x}{\ell} \partial_t - \frac{u x}{\ell} \partial_u + \frac{u^2 \left(\ell^2 + t^2 + x^2 \right) - 1}{2 \ell u^2} \partial_x \ , \\
&& J_{13} = x \partial_t + t \partial_x \ , \\
&& J_{23} = \frac{u^2 \left( - \ell^2 + t^2 + x^2 \right) + 1}{2 \ell u^2} \partial_t - \frac{u t }{\ell} \partial_u + \frac{t x}{\ell} \partial_x \ . 
\end{eqnarray}
It is now straightforward to check that the following definition and subsequent commutators:
\begin{eqnarray}
&& D \equiv J_{02} \ , \quad P \equiv J_{12} - J_{01} \ , \quad K \equiv J_{12} +  J_{01} \ , \\
&& \left[K, P \right] = 2 D \ , \quad \left[ D, P \right] = P \ , \quad \left[ D, K \right] = - K \ . \label{sl2Ralg}
\end{eqnarray}
Evidently, $\{D, K,P\}$ satisfy an $sl(2,R)$ algebra. A general $sl(2,R)$-valued Floquet Hamiltonian can be expanded in this basis:
\begin{eqnarray}
H = \alpha D + \beta P + \gamma K \equiv K^\mu \partial_\mu \ , 
\end{eqnarray}
where $\{\alpha, \beta, \gamma \}$ are three constants. Note that, using the algebra in (\ref{sl2Ralg}), the following adjoint action follows: $D^\dagger = D$, $P^\dagger = K$. Correspondingly, $\alpha \in {\mathbb R}$ and $\beta= \gamma^*$. We will also take $\gamma \in {\mathbb R}$ and therefore $\beta = \gamma$.

The components of the Killing vector $K^\mu$ are given by
\begin{eqnarray}
&& K^t = t \left(\frac{2 \gamma  x}{\ell} - \alpha \right) \ , \quad K^u = u \left(\alpha - \frac{2 \gamma  x}{\ell} \right) \ , \\
&& K^x = \beta  \ell + \frac{\gamma  \left(t^2 - \frac{1}{u^2}\right)}{\ell} + \frac{\gamma  x^2}{\ell} - \alpha  x \ . 
\end{eqnarray}
The Lorentzian norm of this Killing vector can now vanish in the bulk geometry. The corresponding equation yields the Killing horizon:
\begin{eqnarray}
|| K ||^2 = \ell^2 \left[ - u^2 (K^t)^2 + u^2 (K^x)^2 + \frac{1}{u^2}(K^u)^2\right]  = 0 \ , 
\end{eqnarray}
This is a quadratic equation in $u^2$ with a discriminant:
\begin{eqnarray}
{\rm dis} = \frac{\left(\alpha ^2 - 4 \beta  \gamma \right) (\alpha  \ell - 2 \gamma  x)^2}{\ell^2} \ .
\end{eqnarray}
Therefore, real roots ({\it i.e.}~Killing horizons) exist in the regime: $\alpha ^2 - 4 \beta  \gamma \ge 0$ and, in general, there are two distinct real roots. They merge to a single real root when $\alpha ^2 - 4 \beta  \gamma = 0$.\footnote{ Note that, the equation $\alpha  \ell - 2 \gamma  x=0$ can be satisfied for a special value of $x$. At this value, the Killing horizon also becomes an extremal one. } At extremality, when $\alpha ^2 - 4 \beta  \gamma = 0$, the horizon is located at:
\begin{eqnarray}
u_H^2 = \frac{\gamma }{\alpha  \ell x - \gamma  \left(\ell^2 - t^2 + x^2 \right)} = \frac{1}{t^2 - \left( \ell \pm x \right)^2 } \ . \label{extremeuH}
\end{eqnarray}
It can also be checked that the special point at $x=\alpha\ell/(2\gamma)$ yields the following extremal solution: $u_H^2 = 1/t^2$, which is already contained within the solution in (\ref{extremeuH}).

It is now easy to generalize to AdS$_4$, which has the following metric:
\begin{eqnarray}
ds^2 = \ell^2 \left( \frac{du^2}{u^2} + u^2 \left( - dt^2 + dx^2 + dy^2 \right) \right) \ .
\end{eqnarray}
In this case, the ten explicit Killing vectors are given by
\begin{eqnarray}
&& J_{01} = \frac{t x}{\ell} \partial_t - \frac{u x}{\ell} \partial_u + \frac{u^2 \left( - \ell^2 + t^2 + x^2 - y^2 \right) -1}{2\ell u^2} \partial_x + \frac{x y}{\ell} \partial_y \ , \\
&& J_{02} = \frac{t y}{\ell} \partial_ t - \frac{u y }{\ell} \partial_u + \frac{xy}{\ell} \partial_x + \frac{u^2\left(- \ell^2 + t^2 - x^2 + y^2  \right) - 1}{2 \ell u^2} \partial_y \ , \\
&& J_{03} = - t \partial_t + u \partial_u - x \partial_x - y \partial_y \ , \\
&& J_{04} = - \frac{u^2 \left( \ell^2 + t^2 + x^2 + y^2 \right) +1}{2\ell u^2 } \partial_ t + \frac{u t}{\ell} \partial_u - \frac{x t}{\ell} \partial_x - \frac{ t y }{\ell} \partial_ y \ , \\
&& J_{12} = - y \partial_x + x \partial_y \ , \\
&& J_{13} = \frac{t x}{\ell} \partial_t - \frac{u x}{\ell} \partial_u + \frac{u^2\left( \ell^2 + t^2 + x^2 - y^2 \right) - 1}{2 \ell u^2} \partial_x + \frac{x y}{\ell} \partial_y \ , \\
&& J_{14} = x \partial_t + t \partial_x \ , \\
&& J_{23} = \frac{ty}{\ell} \partial_t - \frac{u y}{\ell} \partial_u + \frac{xy}{\ell} \partial_x + \frac{u^2 \left( \ell^2 + t^2 - x^2 + y^2 \right) - 1}{2\ell u^2} \partial_y \ , \\
&& J_{24} = y \partial_t + t \partial_y \ , \\
&& J_{34} = \frac{u^2\left( - \ell^2 + t^2 + x^2 + y^2 \right) +1}{2\ell u^2} \partial_t - \frac{u t}{\ell} \partial_u + \frac{x t}{\ell} \partial_x + \frac{y t}{\ell} \partial_ y \ .
\end{eqnarray}

The Dilatation, Translation and the Special Conformal Transformation generators are identified as:
\begin{eqnarray}
D \equiv J_{03} \ , \quad P_x \equiv J_{13} - J_{01} \ ,  \quad K_x \equiv J_{13}+J_{01} \ .
\end{eqnarray}
The above identifications can be easily verified by noting that 
\begin{eqnarray}
&& \lim_{u \to \infty} J_{13} - J_{01} = \ell \partial_x \ , \\
&& \lim_{u \to \infty} J_{13} + J_{01} = \frac{2 t x}{\ell} \partial_t + \frac{2 x y}{\ell} \partial_y + \frac{t^2 + x^2 - y^2 }{\ell} \partial_x \ ,
\end{eqnarray}
and further the commutation relations:
\begin{eqnarray}
\left[D, P_x \right] =  P_x \ , \quad  \left[ D, K_x\right] = - K_x \ , \quad \left[ K_x, P_x \right] =  2 D \ .
\end{eqnarray}

Given the cotangent basis $\{\partial_t, \partial_u, \partial_x, \partial_y\}$, a general Floquet Hamiltonian $H = \alpha D + \beta P_x + \gamma K_x$ can be written as:
\begin{eqnarray}
&& H = K^\mu \partial_\mu \ , \\
&& K^t = t \left(\frac{2 \gamma  x}{\ell} -\alpha \right) \ , \quad K^u= u \left(\alpha -\frac{2 \gamma  x}{\ell}\right) \ , \\
&& K^x = \beta  \ell + \frac{\gamma  \left(t^2-\frac{1}{u^2}-y^2\right)}{\ell} + \frac{\gamma  x^2}{\ell} - \alpha  x \ , \quad K^y = y \left(\frac{2 \gamma  x}{\ell} - \alpha \right) \ . 
\end{eqnarray}
The Killing horizon can be found by solving the norm of the above Killing vector vanishing with respect to the Poincar\'{e} AdS$_4$ metric. This is given by
\begin{eqnarray}
&& ||K||^2 = 0 \ , \\
&&{\rm disc} = \left(\alpha ^2-4 \beta  \gamma \right) (\alpha  \ell - 2 \gamma  x)^2 \ .
\end{eqnarray}
The discriminant is identical to the AdS$_3$ case hence our conclusions are also identical. The extremal horizon is given by
\begin{eqnarray}
u_H^2 =  \frac{1}{t^2 - \left( \ell \pm x \right)^2 - y^2} \ .
\end{eqnarray}

\begin{small}
	\bibliography{refs}
	\bibliographystyle{jhep}
\end{small}

\end{document}